\documentclass[twocolumn]{aastex631}

\usepackage{amsmath}
\received{3 March 2023}
\revised{3 Nov 2023}
\accepted{12 April 2024}
\submitjournal{ApJ}

\shorttitle{Multi-wavelength Dust Polarization}
\shortauthors{Harrison et al.}
\graphicspath{{./}{figures/}}

\begin{document}

\title{Protoplanetary Disk Polarization at Multiple Wavelengths: Are Dust Populations Diverse?}

\correspondingauthor{Rachel Harrison}
\email{reh3@illinois.edu}

\author[0000-0003-2118-4999]{Rachel E. Harrison}
\affiliation{Department of Astronomy, University of Illinois, Urbana, IL 61801, USA}
\affiliation{School of Physics and Astronomy, Monash University, Vic 3800, Australia}
\author[0000-0001-7233-4171]{Zhe-Yu Daniel Lin}
\affiliation{Department of Astronomy, University of Virginia, Charlottesville, VA 22904, USA}
\author[0000-0002-4540-6587]{Leslie W. Looney}
\affiliation{Department of Astronomy, University of Illinois, Urbana, IL 61801, USA}
\author[0000-0002-7402-6487]{Zhi-Yun Li}
\affiliation{Department of Astronomy, University of Virginia, Charlottesville, VA 22904, USA}
\author[0000-0002-8537-6669]{Haifeng Yang}
\affiliation{Kavli Institute for Astronomy and Astrophysics, Peking University, Yi He Yuan Lu 5, Haidian Qu, Beijing 100871, People’s Republic of China}
\author[0000-0003-3017-4418]{Ian Stephens}
\affiliation{Department of Earth, Environment and Physics, Worcester State University, Worcester, MA 01602, USA}
\author[0000-0001-5811-0454]{Manuel Fern{\'a}ndez-L{\'o}pez}
\affiliation{Instituto Argentino de Radioastronom{\'i}a (CCT-La Plata, CONICET; CICPBA),Buenos Aires, Argentina}

\newcommand{\lwl}[1]{{\color{blue}  lwl; \bf #1}}
\newcommand{\amax}{a_{\text{max}}}


\begin{abstract}

Millimeter and sub-millimeter observations of continuum linear dust polarization provide insight into dust grain growth in protoplanetary disks, which are the progenitors of planetary systems. We present the results of the first survey of dust polarization in protoplanetary disks at 870 $\mu$m and 3 mm. We find that protoplanetary disks in the same molecular cloud at similar evolutionary stages can exhibit different correlations between observing wavelength and polarization morphology and fraction. We explore possible origins for these differences in polarization, including differences in dust populations and protostar properties. For RY Tau and MWC 480, which are consistent with scattering at both wavelengths, we present models of the scattering polarization from several dust grain size distributions. These models aim to reproduce two features of the observational results for these disks: (1) both disks have an observable degree of polarization at both wavelengths and (2) the polarization fraction is higher at 3 mm than at 870 $\mu$m in the centers of the disks. For both disks, these features can be reproduced by a power-law distribution of spherical dust grains with a maximum radius of 200 $\mu$m and high optical depth. In MWC 480, we can also reproduce features (1) and (2) with a model containing large grains ($a_{max}$ = 490 $\mu$m ) near the disk midplane and small grains ($a_{max}$ = 140 $\mu$m) above and below the midplane.

\end{abstract}



\section{Introduction} \label{sec:intro}

Polarized emission from dust on protoplanetary disks has been observed at millimeter and sub-millimeter wavelengths in an increasing number of sources. Multiple mechanisms could theoretically produce this polarized emission, including the alignment of non-spherical dust grains to the magnetic field \citep[e.g.,][]{2007JQSRT.106..225L} or radiation anisotropy \citep{2017ApJ...839...56T}, aerodynamic alignment due to gas-dust interactions \citep[]{1952MNRAS.112..215G, 2019MNRAS.483.2371Y}, or self-scattering of thermal dust emission \citep[e.g.,][]{2015ApJ...809...78K, 2016MNRAS.456.2794Y, 2017MNRAS.472..373Y}. Multi-wavelength polarization observations are necessary to distinguish between these mechanisms and to extract information about dust properties from the polarized emission.

Because magnetic fields are thought to be critical to the accretion process, one of the initial goals of polarization studies of protoplanetary disks was to use polarized emission from grains aligned to the magnetic field to trace the field's morphology \citep[e.g.,][]{2014Natur.514..597S, 2014ApJ...780L...6R, 2015ApJ...798L...2S}. However, with a small number of possible exceptions \citep[e.g.,][]{2018A&A...616A..56A, 2018ApJ...859..165S, 2018ApJ...864...81O}, the polarized emission seen in most disks does not match the pattern that would be produced by dust grains aligning to magnetic field morphologies predicted by theory. Instead, the polarization seen in many disks is better explained by other mechanisms.

In an inclined axisymmetric disk, the collective scattering of thermal photons from dust grains by other dust grains produces polarization that is parallel to the disk minor axis \citep{2015ApJ...809...78K, 2016MNRAS.456.2794Y}. This pattern indicative of scattering has been observed in several sources \citep[e.g., ][]{2017ApJ...851...55S, 2018ApJ...865L..12B, 2019MNRAS.482L..29D}. Scattering polarization fractions of up to a few percent if most of the grains are in the Rayleigh scattering regime (i.e., much smaller than the observing wavelength). For a given observing wavelength, the polarization fraction from scattering is highly dependent on the maximum dust grain size. The polarization fraction peaks when the maximum grain size is of order $\lambda$/2$\pi$, where $\lambda$ is the observing wavelength \citep{2015ApJ...809...78K}. Therefore, observing the scattering polarization spectrum of a source can constrain the maximum grain size. In addition to dust grain size, disk inclination, optical depth, and dust scale height can all affect the level of polarization from scattering \citep[e.g.,][]{2017MNRAS.472..373Y, Ohashi2019, Brunngraber2020}.

Polarization patterns not consistent with pure scattering have also been observed in some protoplanetary disks at millimeter wavelengths. At 3 mm, HL Tau, DG Tau, and Haro 6-13 all exhibit an azimuthal polarization pattern that likely arises from primarily thermal emission by aligned, elongated dust grains \citep[]{2017ApJ...851...55S, 2019ApJ...877L...2H}. Several mechanisms have been proposed for grain alignment, including alignment to the magnetic field or radiation anisotropy through radiative alignment torques (RAT’s) \citep{2017ApJ...839...56T} and mechanical alignment \citep[]{1952MNRAS.112..215G, 2019MNRAS.483.2371Y, 2019ApJ...874L...6K}. HL Tau’s polarization spectrum is particularly interesting: the disk’s polarization morphology transitions from a pattern consistent with scattering at $\sim$870 $\mu$m to an elliptical pattern at $\sim$3 mm, with an intermediate pattern at $\sim$1.3 mm \citep{2017ApJ...851...55S}. Recently, \citet{2022MNRAS.512.3922L} and \citet{2024MNRAS.528..843L} demonstrated that HL Tau’s polarization spectrum could be explained by polarization from thermal emission and scattering of aligned grains, and \citet{2014Natur.514..597S} demonstrated that the polarization seen at high resolution in HL Tau at 870 $\mu$m could be explained by the same mechanisms. 
At longer wavelengths, thermal polarization dominates because the low optical depth makes scattering inefficient. At shorter wavelengths, scattering polarization dominates because the high optical depth makes scattering events frequent and dichroic extinction decreases the contribution from thermal polarization. \citep{2022MNRAS.512.3922L}.

The changes in polarization with wavelength seen in HL Tau invite the question of whether other disks exhibit similar polarization patterns at these wavelengths. \citet{2019ApJ...877L...2H} demonstrated that otherwise similar protoplanetary disks can exhibit different polarization morphologies at the same observing wavelength. The polarized emission in the Class II protoplanetary disks observed at 3 mm with ALMA in \citet{2019ApJ...877L...2H} can be qualitatively split into two categories: those with polarization consistent with scattering (RY Tau, MWC 480, and potentially DL Tau), and those with azimuthally-oriented polarization vectors (DG Tau, Haro 6-13, and V892 Tau). 
Given how strongly scattering is affected by the grain size with respect to the observing wavelength and how thermal polarization from aligned grains depends on the optical depth, differences in the polarization spectra could indicate differences in the grain sizes and conditions of grain alignment. These dust grains are the building blocks of potential planets in such systems. We present the results of total intensity and continuum linear polarization observations of five Class II disks at 870 $\mu$m and 3 mm: Haro 6-13, MWC 480, RY Tau, DL Tau, and V892 Tau. With these observations, we have significantly expanded the number of protoplanetary disks observed in polarization at multiple wavelengths. 

\section{Observations}
We observed DL Tau, Haro 6-13, RY Tau, MWC 480, and V892 Tau at 870 $\mu$m (ALMA Band 7). These sources are all Class II disks located in the Taurus Molecular Cloud. The disks surround low-mass protostars with a range of ages, luminosities, and masses. The disks of DL Tau, Haro 6-13, RY Tau, and MWC 480 all surround single stars, while V892 Tau is a circumbinary disk around a close binary with a third stellar object orbiting outside the disk \citep{2021ApJ...915..131L}. Protostar masses, luminosities, ages, disk inclinations, and distances are listed in Table \ref{tab:disk_params}. Table \ref{ring_gap_radii} lists the radii of known rings and gaps in the disks. 

\begin{deluxetable*}{ccccccc}[t]




\tablecaption{Disk parameters}

\tablenum{1}

\tablehead{\colhead{Source} & \colhead{$M_*$} & \colhead{$L_*$} & \colhead{Age} & \colhead{Inc.} & \colhead{Dist.} & \colhead{Citation} \\ 
\colhead{} & \colhead{($M_\odot$)} & \colhead{($L_\odot$)} & \colhead{(Myr)} & \colhead{(deg.)} & \colhead{(pc)} & \colhead{} } 

\startdata
Haro 6-13 &  1.00$\pm$0.15 & 0.79 & 2.60 &  38 &  130.4 & 1, 2, 3 \\
MWC 480 &  1.91 & 17.38 & 6.90 & 36.5 &  161 & 4\\
RY Tau &  2.04 & 12.30 & 5.00 & 65.0 &  128 & 4\\
V892 Tau &  6.0$\pm$0.2\tablenotemark{a} & 128.82 & 0.79 & 54.5 &  134.5 & 5, 6\\
DL Tau &  0.98 & 0.65 & 3.50 & 45.0 &  159 & 4\\
\enddata
\tablenotetext{a}{Combined mass of binary.}

\tablecomments{Masses, inclinations and distances for the sources in this survey. A disk inclination of 0$^\circ$ is defined as face-on, and an inclination of 90$^\circ$ is defined as edge-on. Distances are from Gaia parallax measurements \citep{Gaia2018, Gaia2016}. \\
Citations: 1: \citet{2009ApJ...701..698S}, 2: \citet{Garufi2021}, 3: \citet{2019ApJ...882...49L}, 4: \citet{2018ApJ...869...17L}, 5: \citet{2021ApJ...915..131L}, 6: \citet{2014ApJ...786...97H}.}

\label{tab:disk_params}
\end{deluxetable*}

\begin{deluxetable*}{ccccccc}
\tablecaption{Ring and gap radii, in au}
\tablenum{2}
\tablehead{\colhead{Source} & \colhead{Ring 1} & \colhead{Ring 2} & \colhead{Ring 3} & \colhead{Gap 1} & \colhead{Gap 2} & \colhead {Gap 3} } 
\startdata
MWC 480 & 97.58$\pm$0.08 & - & - & 73.43$\pm$0.16 & - & -\\
RY Tau & 18.19$\pm$0.00 & 49.04$\pm$0.14 & - & 43.41$\pm$0.13 & - & -\\
V892 Tau & 26.90$\pm$0.14 & - & - & - & - & -\\
DL Tau & 46.44$\pm$0.48 & 78.08$\pm$0.24 & 112.27$\pm$0.32 & 39.29$\pm$0.32 & 66.95$\pm$0.87 & 88.90$\pm$0.11 \\
\enddata
\tablecomments{Radii of known rings and gaps, assuming the distances in Table \ref{tab:disk_params}. Data for MWC 480, RY Tau, and DL Tau from \citet{2018ApJ...869...17L}. Data for V892 Tau from \citet{2021ApJ...915..131L}.}
\label{ring_gap_radii}
\end{deluxetable*}

The observations were taken with ALMA between 18 September and 19 September 2018 in configuration C43-4. The observations were at a frequency range of 337.48-349.51 GHz (ALMA Band 7). J0438+3004 was the phase calibrator, J0522--3627 was the polarization calibrator, and J0510+1800 was the bandpass and flux calibrator.

The datasets were calibrated by data analysts at the North American ALMA Science Center. After this initial calibration, we performed three rounds of phase-only self-calibration on all Stokes parameters ($I$, $Q$, $U$, and $V$). The gain calibration solution interval was equal to the scan length for the first round of self-calibration, 30 seconds for the second round, and 15 seconds for the third round. We used the CASA task tclean with stokes='IQUV', Briggs weighting, and a robust parameter of 0.5. Polarization angle and intensity maps were produced from the Stokes $Q$ and $U$ data. The polarized intensity maps were debiased using the average noise value determined from the $Q$ and $U$ maps, an estimator used by e.g. \citet{1974ApJ...194..249W} and \citet{2016MNRAS.461..698V}: 

\begin{equation}
  P=\begin{cases}
    \sqrt{Q^2 + U^2 - \sigma^2} & \text{if $\sqrt{Q^2 + U^2} \geq \sigma$} \\
    0 & \text{otherwise}
  \end{cases}
\end{equation}

While V892 Tau and DL Tau were reported as unpolarized at 3 mm in \citet{2019ApJ...877L...2H}, upon closer examination we found that small regions of these disks were polarized at the $>3\sigma$ level. V892 Tau exhibits azimuthal polarization at 3 mm. 

DL Tau has a polarized region near the center of the disk that is less than half the size of the beam, and the peak polarized intensity is $3.5\sigma_P$. This region's peak polarization fraction is 0.9\%, and the direction of polarization is along the disk's minor axis. DL Tau also has several smaller polarized regions outside the disk center with polarization fractions up to 30\%, whose direction of polarization angle is along the disk's minor axis. The polarized regions in DL Tau are so small that few independent measurements of the polarization angle can be made. Higher sensitivity observations will be needed to determine whether this tentative polarization detection is real. We also note that while the majority of the polarized emission at Band 3 in MWC 480 is consistent with scattering, with the angle of polarization along the disk minor axis, there are two sub-beam-sized regions of polarized emission $>3\sigma$ in the outer part of the disk whose polarization angles are not along the minor axis. It is possible that these non-scattering polarized regions arise from aligned grains; however, 
they do not form a complete ring as the azimuthally-oriented polarized emission in DG Tau, Haro 6-13, and HL Tau does at 3 mm. 

The images had angular resolutions between 0.2 and 0.3 arcseconds. The uncertainty on absolute flux calibrations with ALMA is estimated at $\sim$10\%. ALMA's instrumental limit for a 3$\sigma$ detection of polarized emission is 0.1\% polarization for compact sources within one-third of the primary beam. The typical sensitivities of the Stokes $Q$ and $U$ images are about 50-60 $\mu$Jy/beam. For the rest of this paper only statistical uncertainties are considered. 

\begin{deluxetable*}{cccccccc}
\tablecaption{Total and polarized intensities and RMS values, peak percent polarization, and beam sizes}

\tablenum{3}

\tablehead{\colhead{Source} & \colhead{$I$ flux} & \colhead{$I$ peak} & \colhead{$\sigma_I$} & \colhead{$P$ peak} & \colhead{$\sigma_P$} & \colhead{Peak \%$P$} & \colhead{Beam FWHM}\\ 
\colhead{} & \colhead{(mJy)} & \colhead{(mJy bm$^{-1}$)} & \colhead{($\mu$Jy bm$^{-1}$)} & \colhead{($\mu$Jy bm$^{-1}$)} & \colhead{($\mu$Jy bm$^{-1}$)} & \colhead{} } 

\startdata
Haro 6-13 & 339 $\pm$ 1 & 130.2 & 102 & 1074 & 61.6 & 1.9 & $0\farcs230 \times 0\farcs172$\\ 
MWC 480 & 598 $\pm$ 16 & 131.7 & 176 & 540 & 54.8 & 1.6 & $0\farcs254 \times 0\farcs170$\\
RY Tau & 570 $\pm$ 6 & 111.9 & 98.4 & 1197 & 58.1 & 3.2 & $0\farcs248 \times 0\farcs174$\\
V892 Tau & 745 $\pm$ 44 & 137.5 & 152 & 968 & 59.6 & 2.0 & $0\farcs244 \times 0\farcs172$\\
DL Tau & 310 $\pm$ 17 & 53.9 & 98.9 & 340 & 59.2 & 1.3 & $0\farcs235 \times 0\farcs173$\\
\label{obs_870}
\enddata



\end{deluxetable*}

\section{Results} \label{sec:results}

\begin{figure*}[ht]
\gridline{\fig{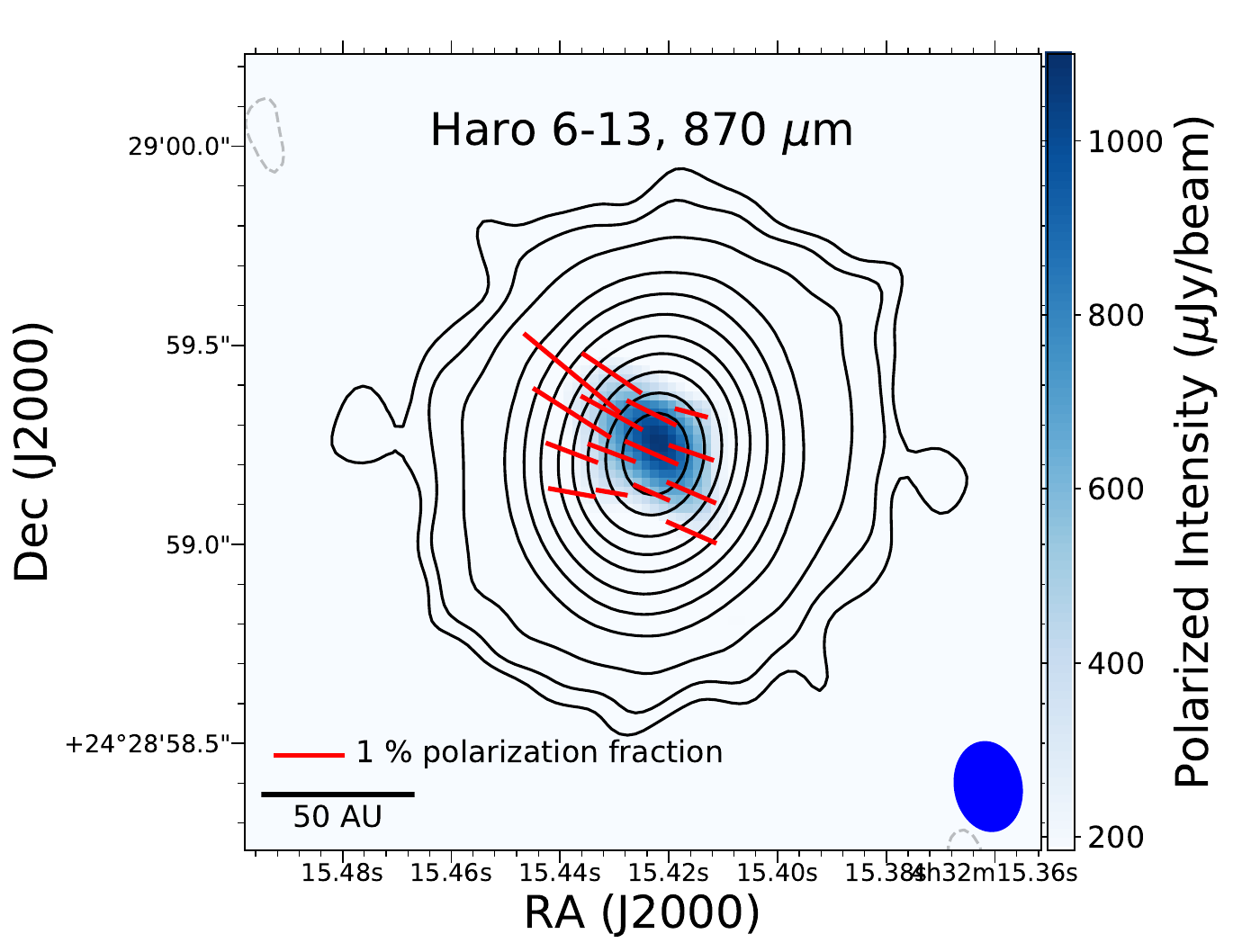}{0.3\textwidth}{(a)}
          \fig{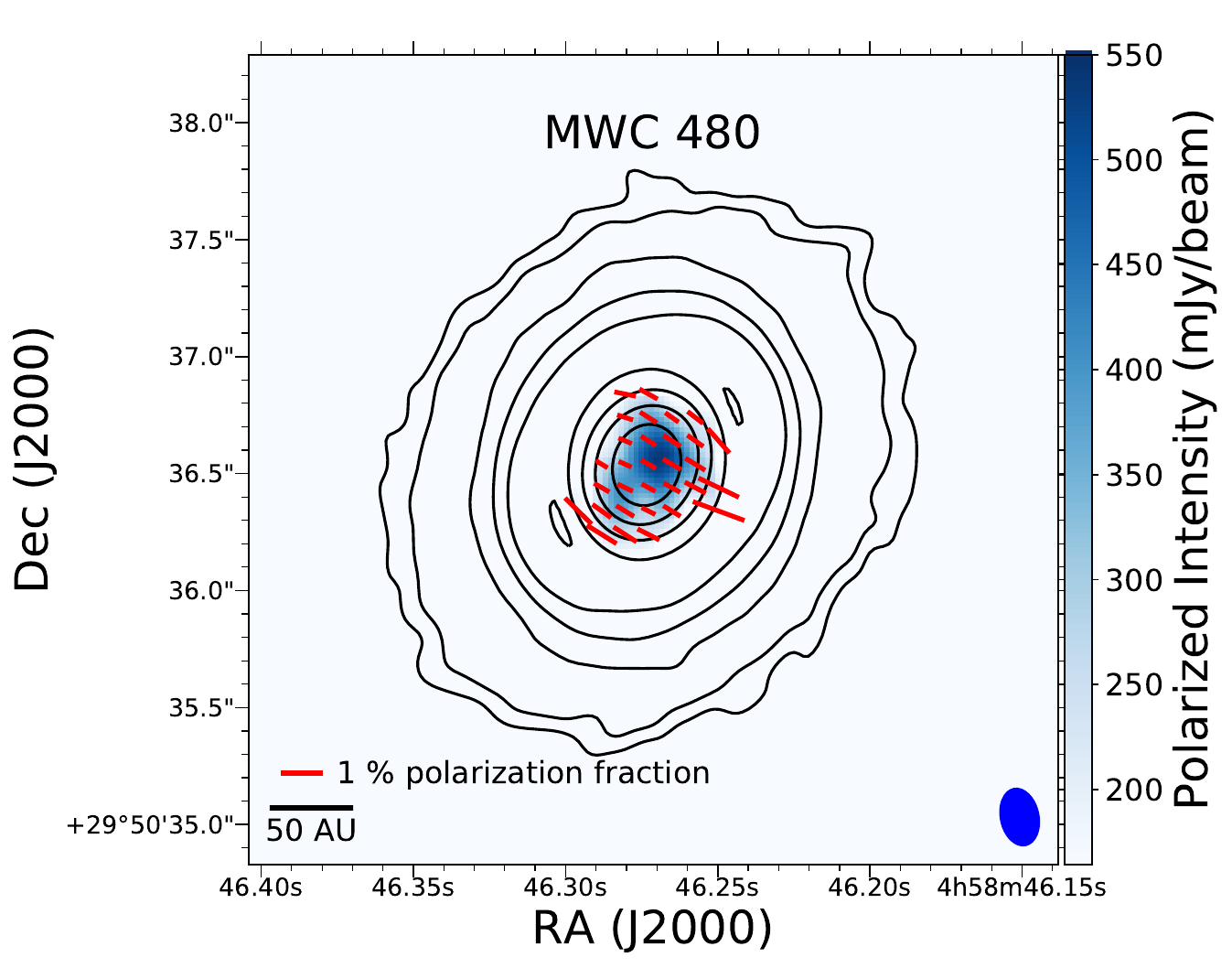}{0.3\textwidth}{(b)}
          \fig{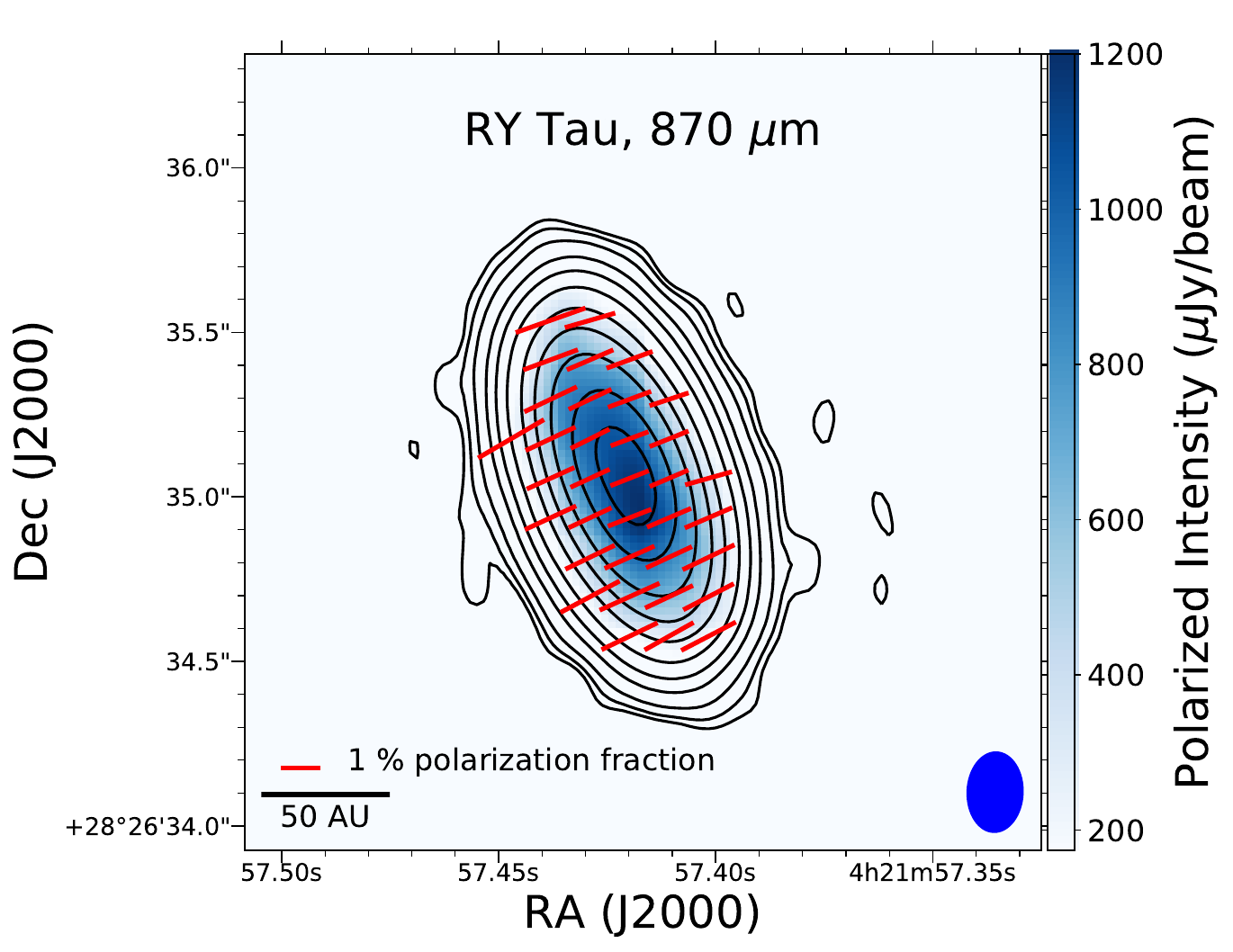}{0.3\textwidth}{(c)}}
\gridline{\fig{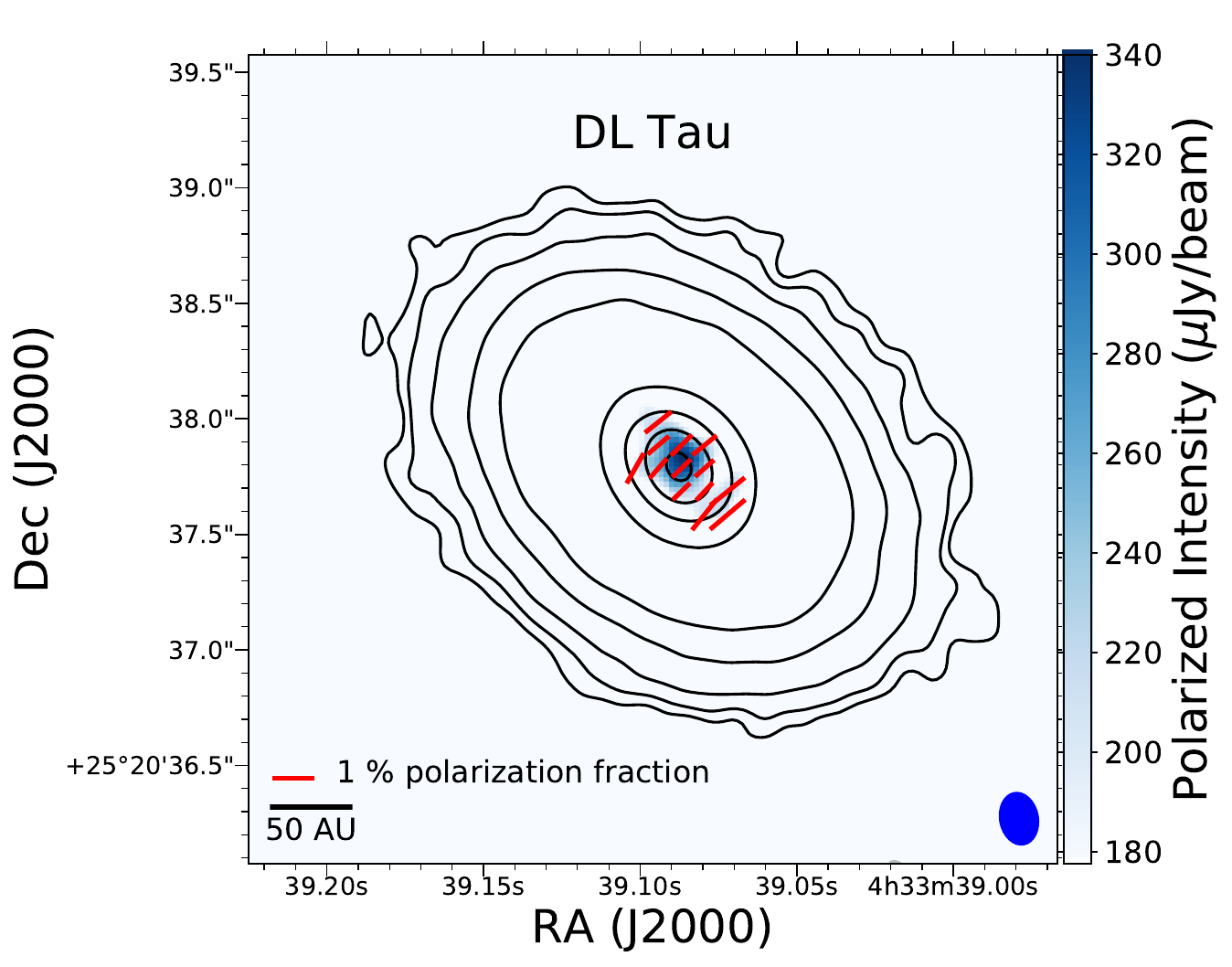}{0.3\textwidth}{(d)}
          \fig{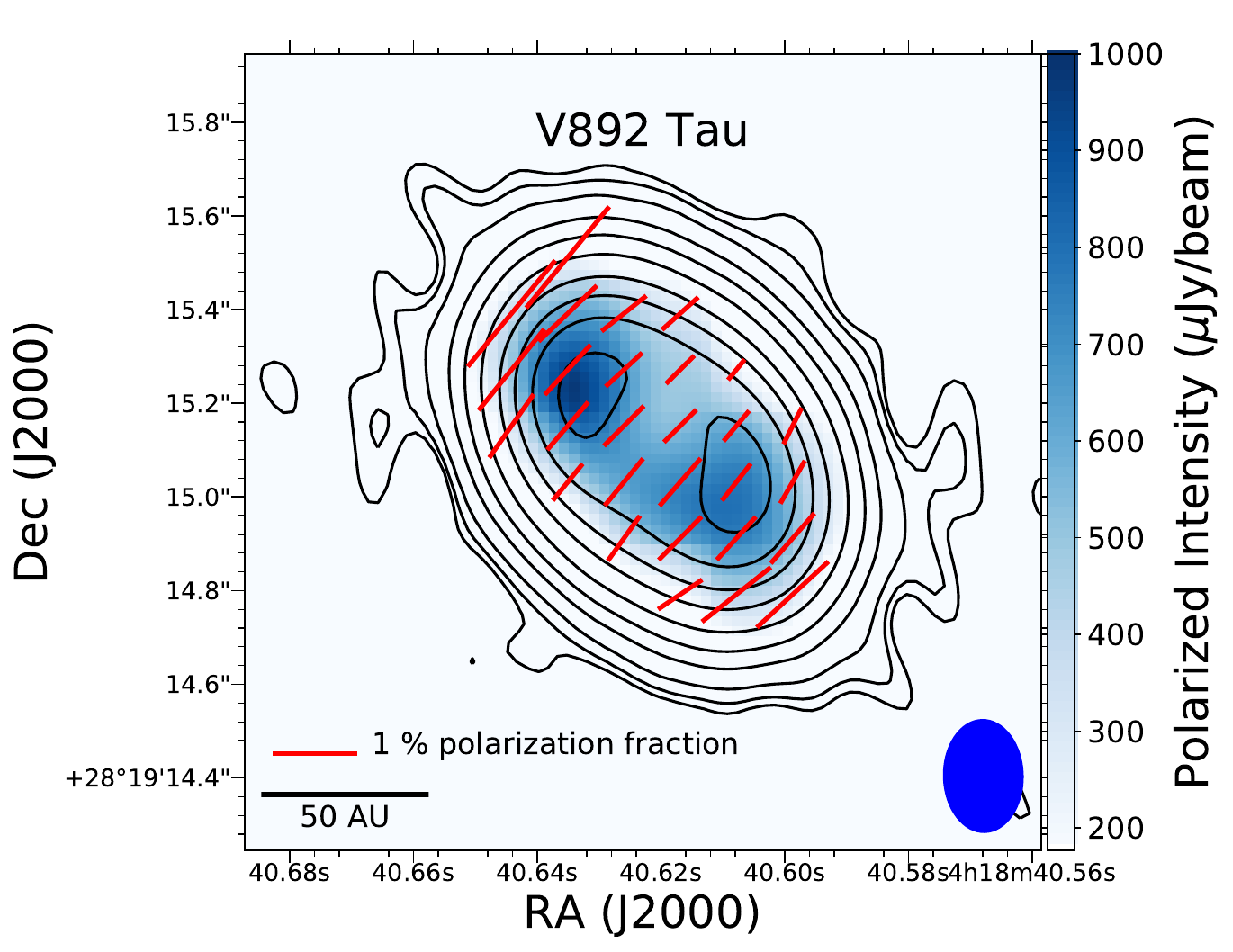}{0.3\textwidth}{(e)}}

\caption{Images of Haro 6-13, MWC 480, RY Tau, DL Tau, and V892 Tau at 870 $\mu$m. The contours represent total intensity (Stokes $I$) of -3 (dashed), 3, 10, 50, 100, 250, 325, 500, 750, 1000, and 1500$\sigma$ levels, where $\sigma$ is listed in Table \ref{obs_870}. The colormap represents polarized intensity with the scale on the right of each source, and is shown where the polarized intensity is $>3\sigma_P$. The length of the polarization vectors corresponds to the polarization fraction. The vectors are plotted with $\sim$ 3 segments per beam, and are plotted in regions where $P$ and $I$ are both $>3\sigma$.}
\label{maps2}
\end{figure*}

Here, we describe the results of the 870 $\mu$m observations and compare them to the polarized emission observed in the same sources at 3 mm in \citet{2019ApJ...877L...2H}. At 870 $\mu$m, all five disks exhibit polarization morphologies consistent with scattering, regardless of stellar mass, age, luminosity, multiplicity, or polarization at 3 mm. Figure \ref{maps2} shows the total intensity, polarized intensity, polarization angle, and polarization fraction for the five disks at 870 $\mu$m. Other observations of protoplanetary disks at 870 $\mu$m \citep[e.g., ][]{2017ApJ...851...55S, 2018ApJ...865L..12B, 2019MNRAS.482L..29D} and 1.3 mm \citep{Sadavoy2019} have found that scattering polarization is fairly common at these wavelengths. This indicates that dust in these sources has grown large enough to produce scattering polarization at wavelengths of $\sim$1 mm (likely a few hundred $\mu$m, assuming the grains are compact and spherical). These disks must also be optically thick enough ($\tau\gtrsim1$) at the observed wavelengths to produce an observable polarization fraction.


Table \ref{obs_870} lists the measured values for total intensity (as determined from a Gaussian fit to the source) and peak polarized intensity, as well as the beam size. Figure \ref{maps2} shows composite images of the Stokes $I$ and polarized emission observed in Haro 6-13, MWC 480, RY Tau, DL Tau, and V892 Tau at 870 $\mu$m. The contours indicate total intensity of the dust emission in multiples of $\sigma$, and the blue shading indicates the polarized intensity. The red lines indicate the polarization angle, and are scaled to correspond to the polarization fraction. We refer to these lines as ``vectors'', though there is a 180$^\circ$ ambiguity in their direction. 

\begin{figure*}[ht]
\gridline{\fig{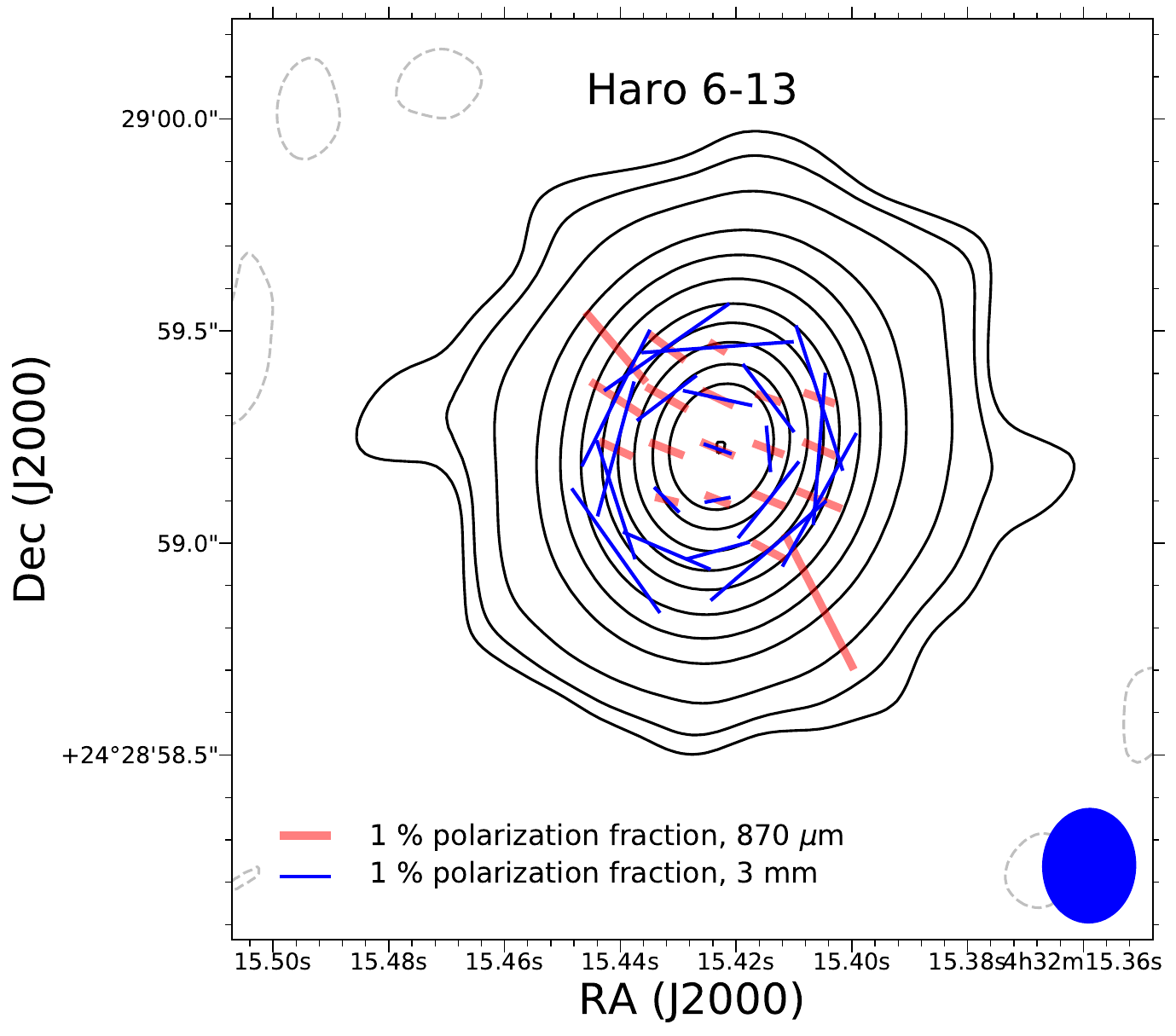}{0.4\textwidth}{(a)}
          \fig{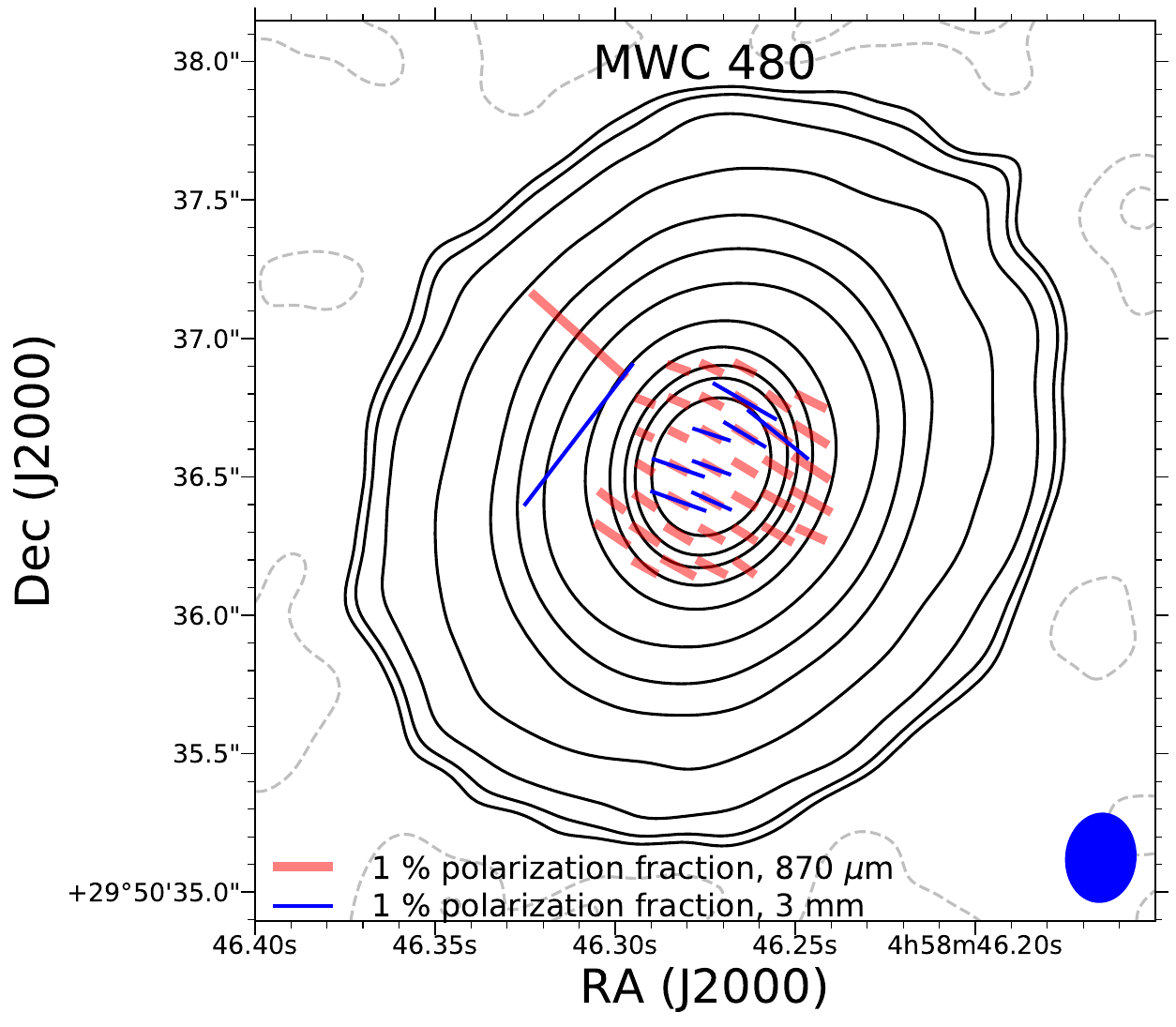}{0.4\textwidth}{(b)}}
\gridline{\fig{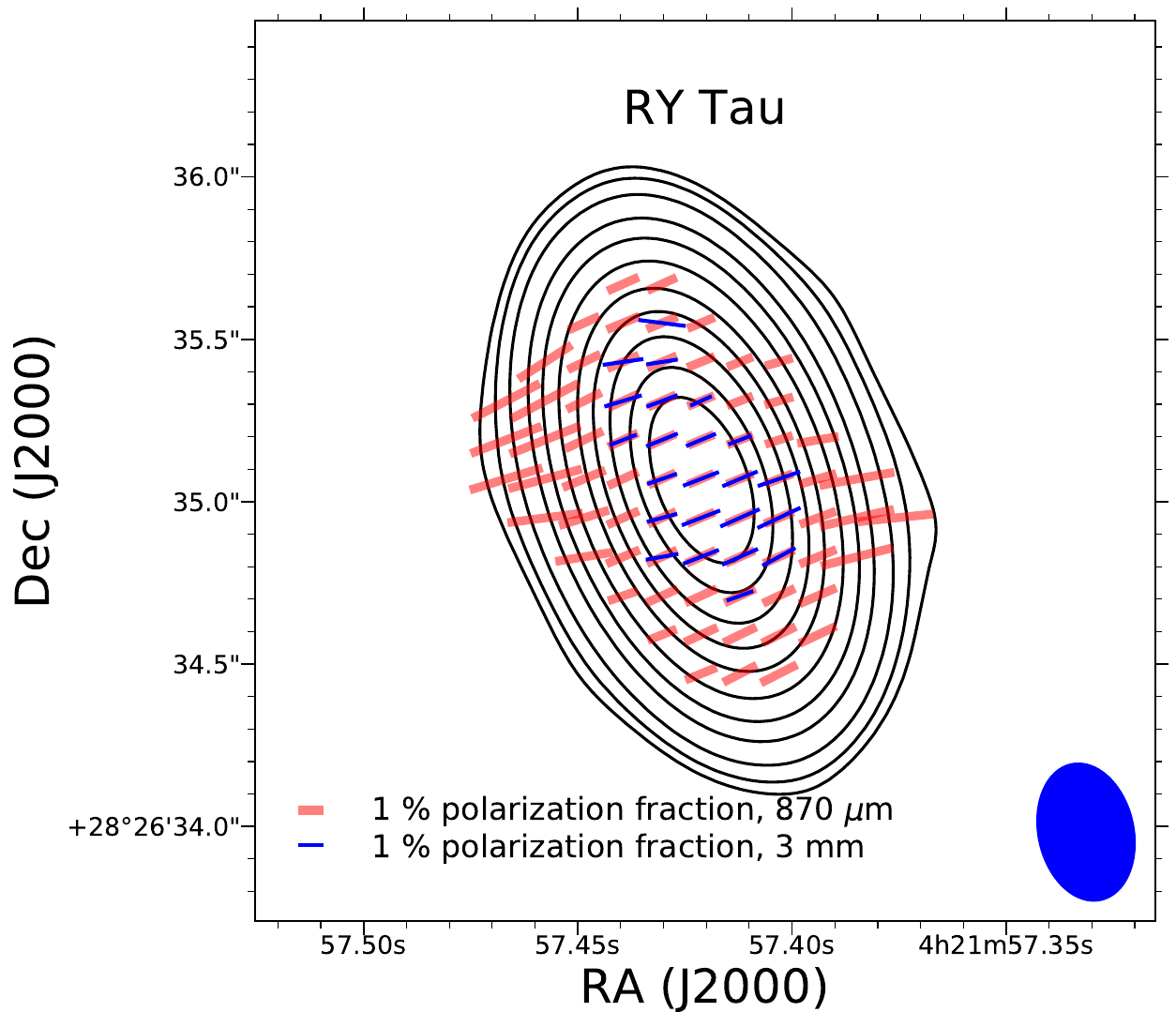}{0.4\textwidth}{c)}
          \fig{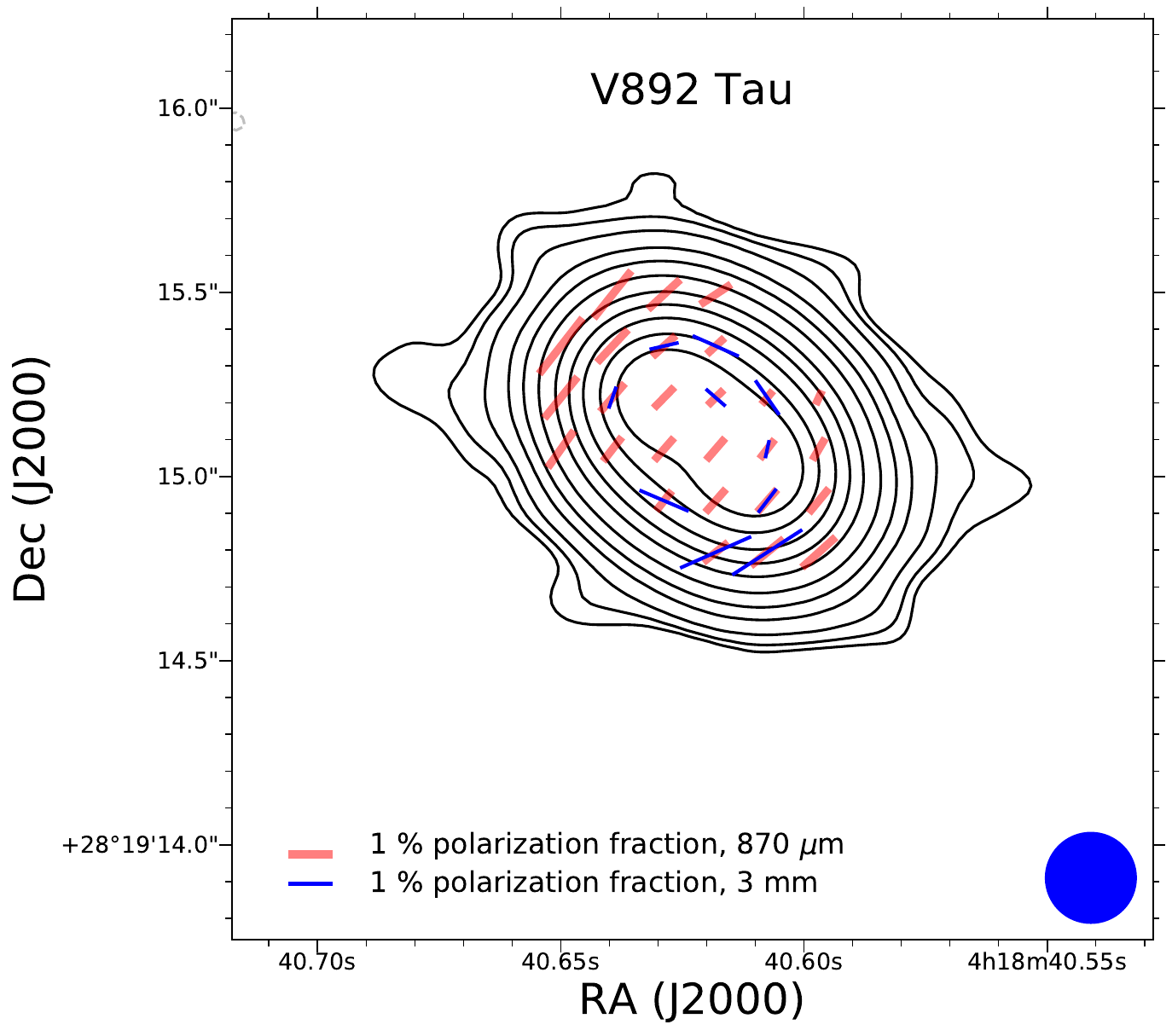}{0.4\textwidth}{(d)}}
\gridline{\fig{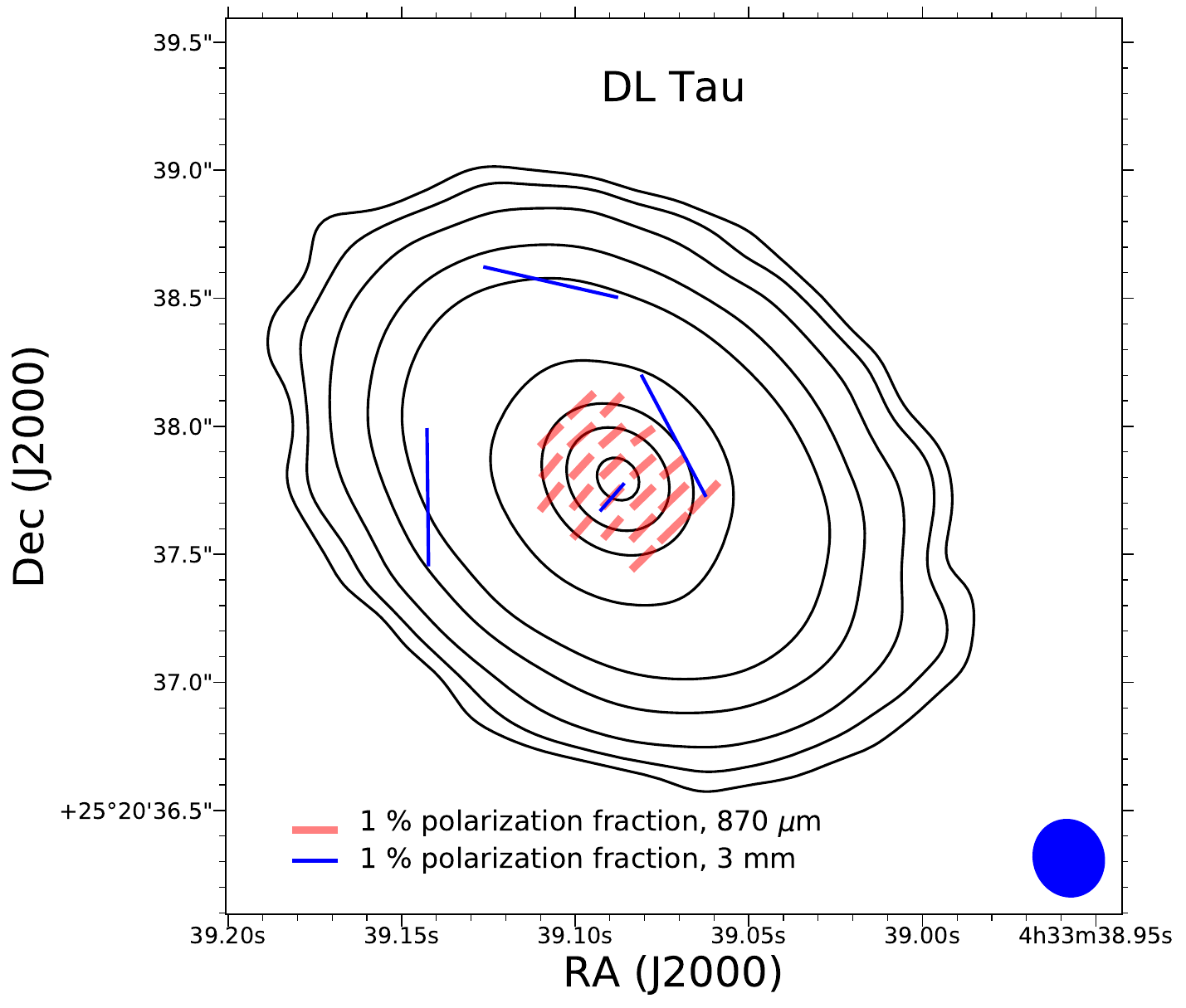}{0.4\textwidth}{e)}}
\caption{Images of Haro 6-13, MWC 480, RY Tau, V892 Tau, and DL Tau showing the polarization angles at 870 $\mu$m (red lines) and 3 mm (blue lines). The black contours represent Stokes $I$ at levels of -3 (dashed), 3, 10, 50, 100, 250, 325, 500, 750, 1000, and 1500$\sigma$, where $\sigma$ is the RMS of the smoothed 870 $\mu$m images. Vectors are plotted in regions where the total intensity and polarized intensity are both $>3\sigma$. The length of the vectors is scaled with percent polarization up to a threshold of 3\% to better show the variation at low polarization fraction. The blue ellipse represents the beam size.}
\label{twoband_composite}
\end{figure*}

Figure \ref{twoband_composite} shows the polarization fractions and angles for each disk detected in polarization at both 3 mm and 870 $\mu$m. 
The vectors are superimposed on the Stokes $I$ contours at 870 $\mu$m from Figure \ref{maps2}. The data in Figure \ref{twoband_composite} have been smoothed to the same resolution. For Haro 6-13, MWC 480, and RY Tau, the 870 $\mu$m data have been smoothed to the resolution of the 3 mm data. The negative bowl seen around MWC 480 is also present in the non-smoothed data, and may be due to large-scale structure being resolved out. The V892 Tau data from both bands has been smoothed to a resolution of $0 \farcs 25$; this resolution is slightly larger than the major and minor axes of the beams at 3 mm and 870 $\mu$m. The angular resolutions of the data in Figure \ref{twoband_composite} are listed in Table \ref{tab:3mm_res}.

Polarized emission was not detected in the center of V892 Tau’s disk at 3 mm. We examined the possibility that non-thermal emission could affect the observed polarization fraction. Since only thermal emission contributes to polarized emission by scattering, contamination from non-thermal emission could reduce the observed polarization fraction. \citet{2021ApJ...915..131L} observed a spectral index of $\sim$0 in V892 Tau between 8 mm and 9.8 mm near the center of the disk, which they report may be due to free-free emission from ionized gas. \citet{Hamidouche2010} estimated the contamination from free-free emission in V892 Tau to be 10\% at 2.7 mm. If the level of contamination is similar at 3 mm, then the free-free emission would only lower the observed polarization fraction by $\sim$0.1\%, which is at the threshold of ALMA’s polarization detection limits. If the azimuthally-oriented polarization pattern continues in the center of V892 Tau, beam smearing effects are a more likely explanation for the lack of polarized emission in the disk’s center.


\begin{deluxetable}{cc}[t]
\tablecaption{Angular resolutions of smoothed data}
\tablenum{4}
\tablehead{\colhead{Source} & \colhead{Beam (3 mm)} } 
\startdata
Haro 6-13 & 0$\farcs$27$\times$0$\farcs$22\\
MWC 480 & 0$\farcs$33$\times$0$\farcs$26\\
RY Tau & 0$\farcs$43$\times$0$\farcs$30\\
V892 Tau & 0$\farcs$25$\times$0$\farcs$25\\
DL Tau & 0$\farcs$31$\times$0$\farcs$28\\
\label{tab:3mm_res}
\enddata

\end{deluxetable}

\begin{figure*}[ht]
\gridline{\fig{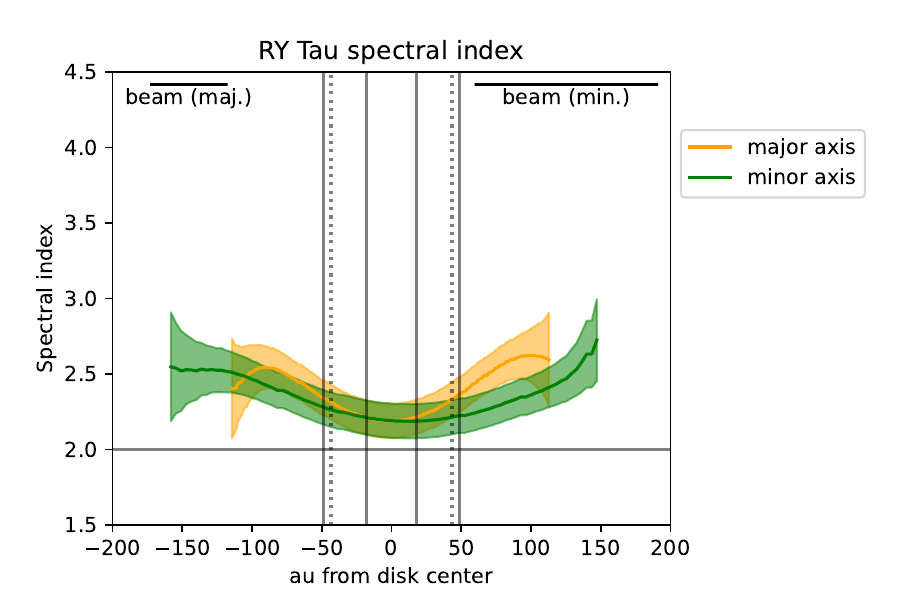}{0.48\textwidth}{(a)}
          \fig{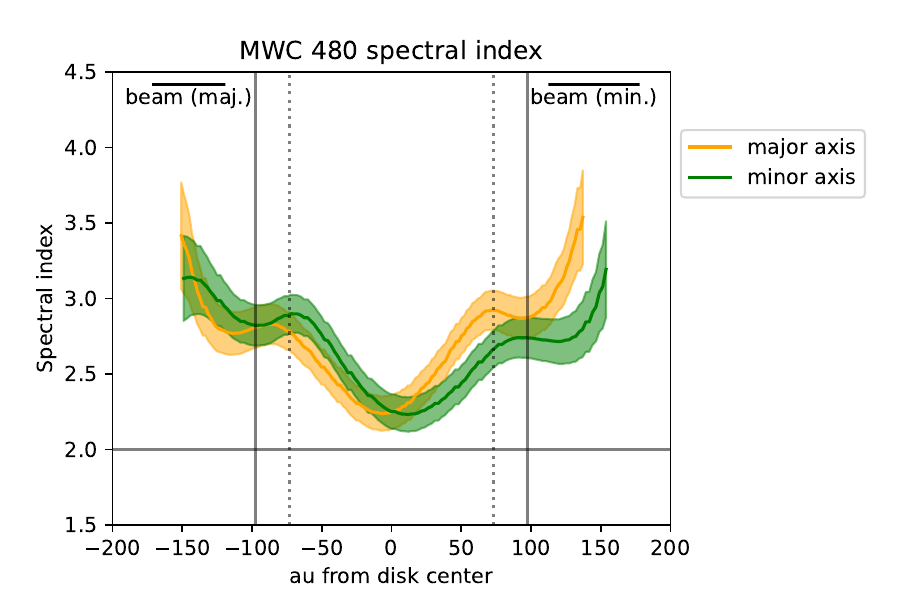}{0.48\textwidth}{(b)}}
\gridline{\fig{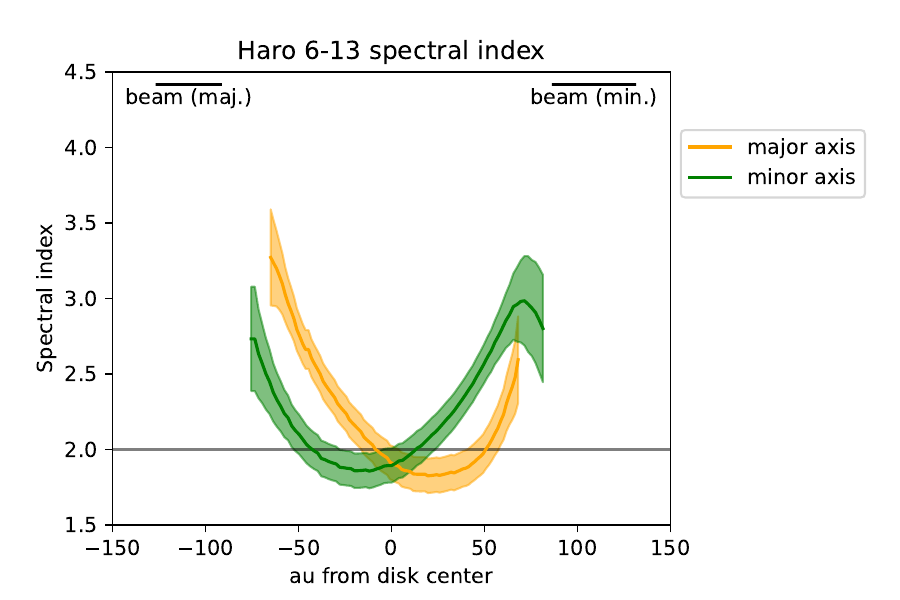}{0.48\textwidth}{(c)}
          \fig{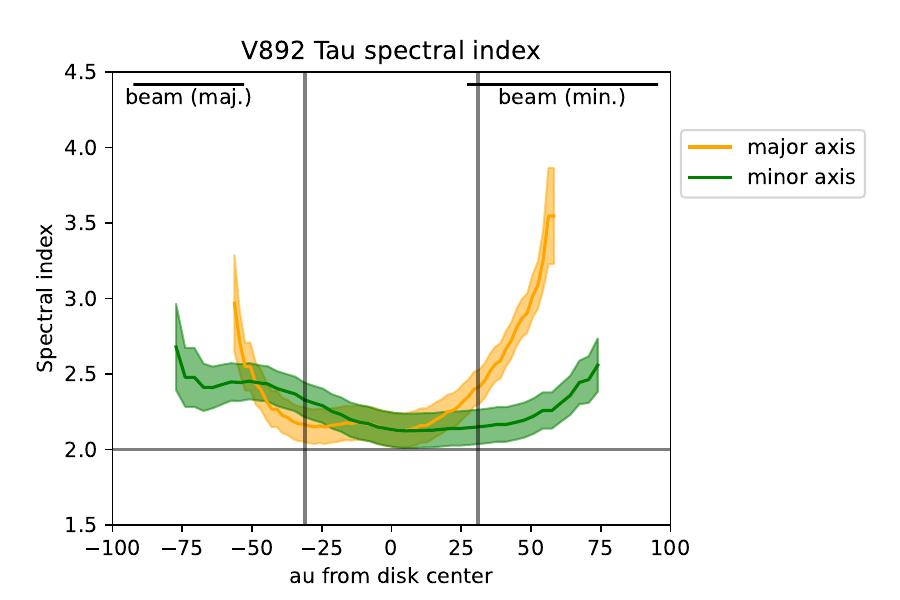}{0.48\textwidth}{(d)}}

\caption{Spectral indices along the disks' major and minor axes. Shaded region represents 1$\sigma$ error bars, including the 10\% uncertainty on the absolute flux value. The gray horizontal line represents $\alpha = 2$.}
\label{spix}
\end{figure*}

We calculated the average values of the spectral index ($\alpha$) in a region at the centers of the disks with an area equivalent to the beam FWHM in Table \ref{tab:3mm_res}. The average values of $\alpha$ at the centers of the disks were 2.0 for Haro 6-13, 2.5 for MWC 480, 2.3 for RY Tau, and 2.3 for V892 Tau. There was no obvious correlation between the spectral indices at the centers of the disks and their polarization spectra. These values of $\alpha$ are consistent with the disk either being optically thick in the center or having small $\beta$. Figure \ref{spix} shows the radial profiles of the spectral index along the disks' major and deprojected minor axes. Full optical depth modeling of these disks would require high-resolution observations that resolved disk substructure, namely rings and gaps, since disk substructure can significantly affect both the spectral index and polarization fraction and morphology. 

\begin{figure*}[ht]
\gridline{\fig{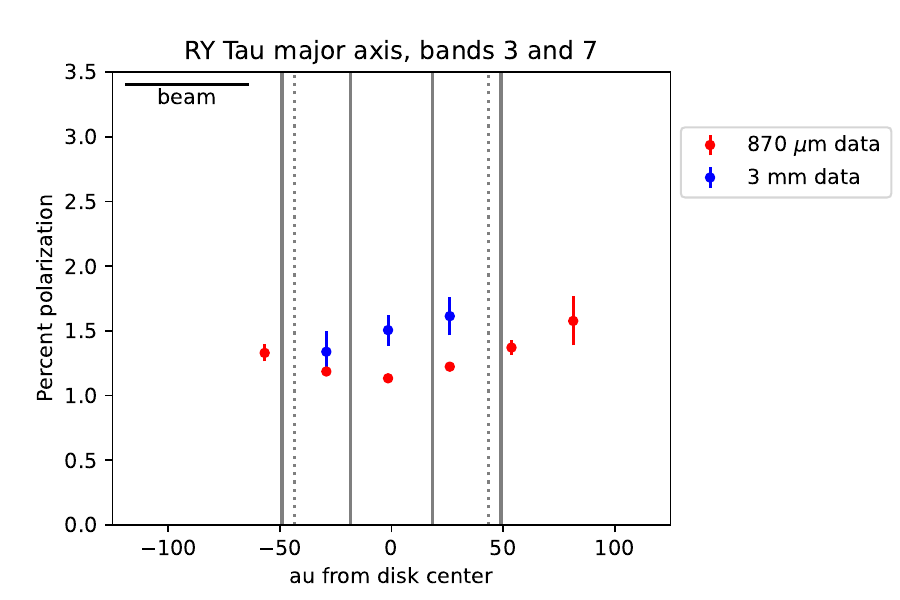}{0.48\textwidth}{(a)}
          \fig{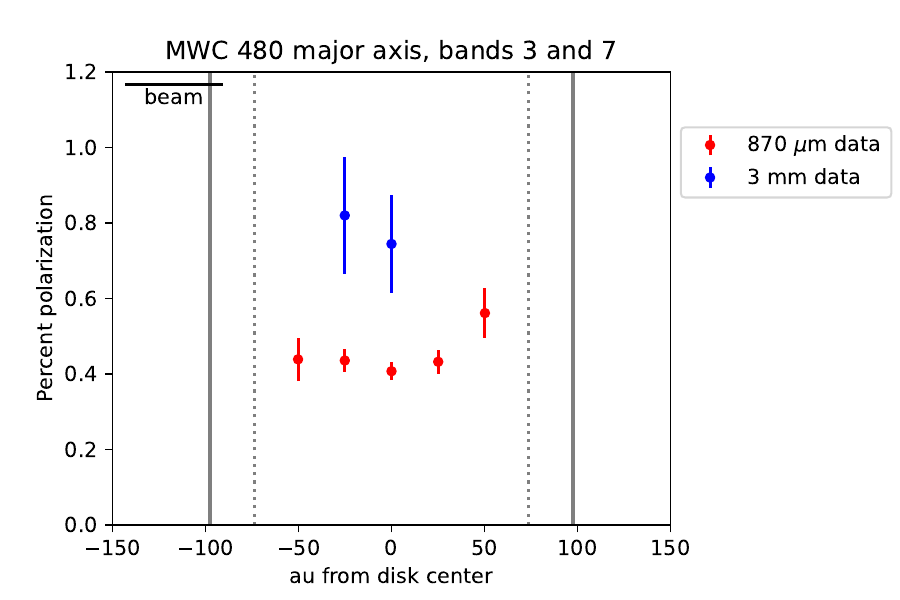}{0.48\textwidth}{(b)}}
\gridline{\fig{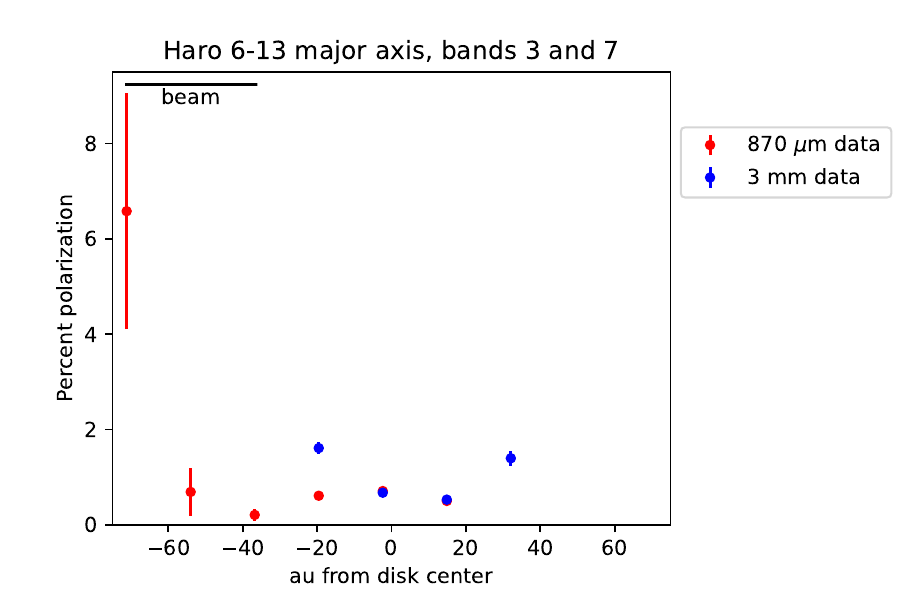}{0.48\textwidth}{(c)}
          \fig{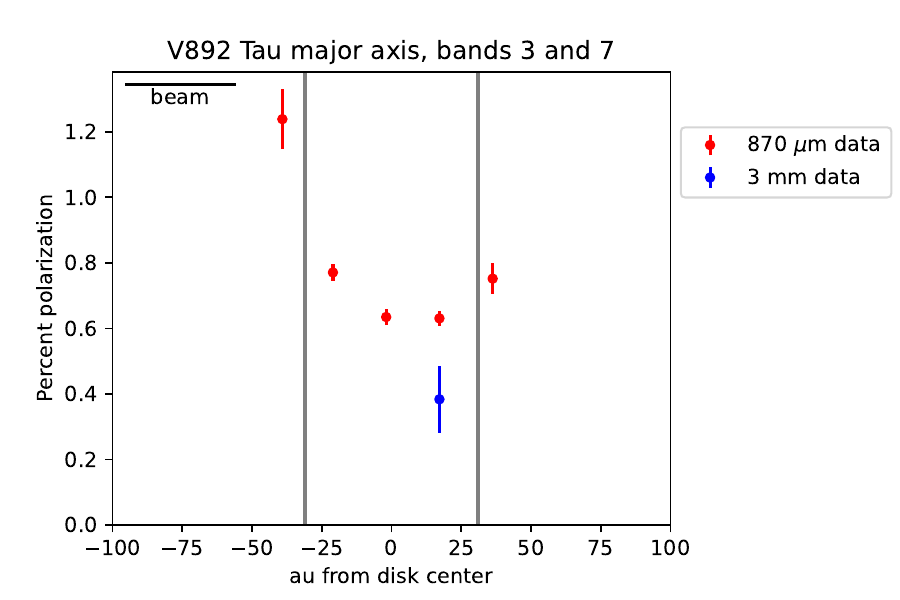}{0.48\textwidth}{(d)}}
\gridline{\fig{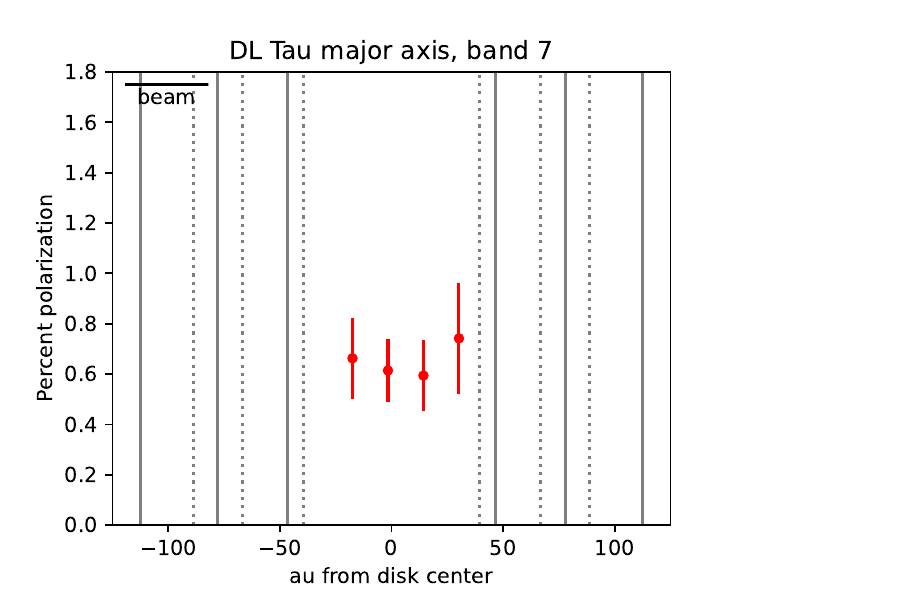}{0.48\textwidth}{(e)}
          \fig{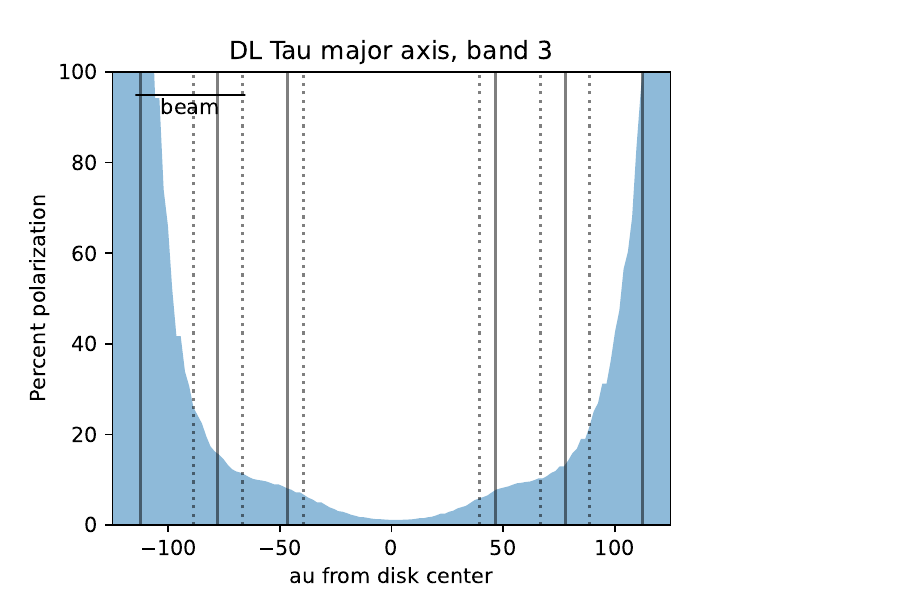}{0.48\textwidth}{(d)}}

\caption{Percent polarization vs. distance from disk center along the disk major axis at 3 mm and 870 $\mu$m for RY Tau, MWC 480, Haro 6-13, and V892 Tau, and at 870 $\mu$m for DL Tau. For RY Tau, MWC 480, Haro 6-13, and V892 Tau, the two bands have been smoothed to the same resolution; the scale bar indicates the beam size. The smoothed resolutions are listed in Table \ref{tab:3mm_res}. For DL Tau, we show the 870 $\mu$m data at the resolution listed in Table \ref{obs_870}. Since the polarized region in DL Tau at 3 mm was less than half of the beam FWHM across, we have plotted the 3$\sigma$ upper limit on the percent polarization in DL Tau at 3 mm. The resolution at 3 mm is listed in Table \ref{tab:3mm_res}.} The data are only plotted in regions where the polarized intensity is at least 3$\sigma_P$. Solid vertical lines indicate the location of known rings, and dotted vertical lines indicate the location of known gaps.
\label{twoband_pfrac_majax}
\end{figure*}

\begin{figure*}[ht]
\gridline{\fig{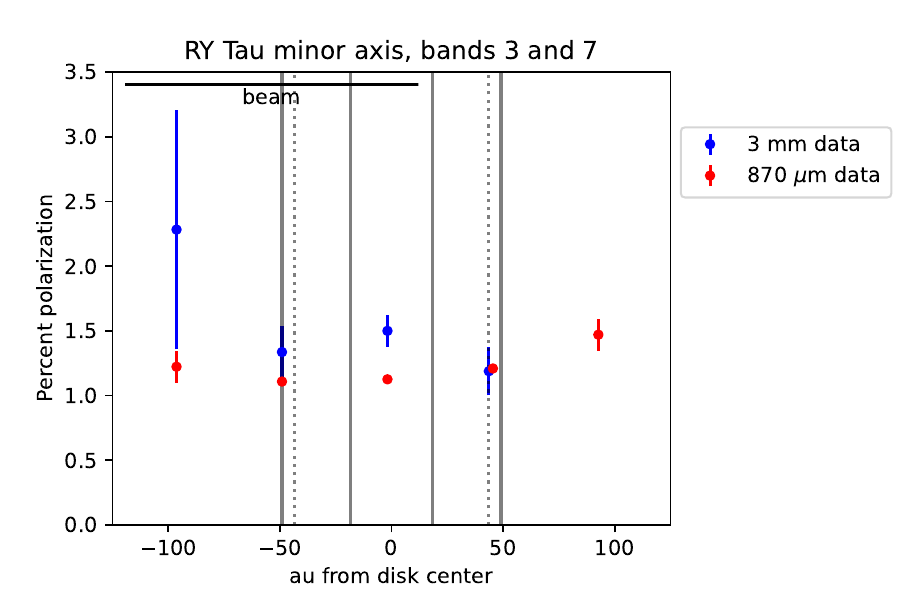}{0.48\textwidth}{(a)}
          \fig{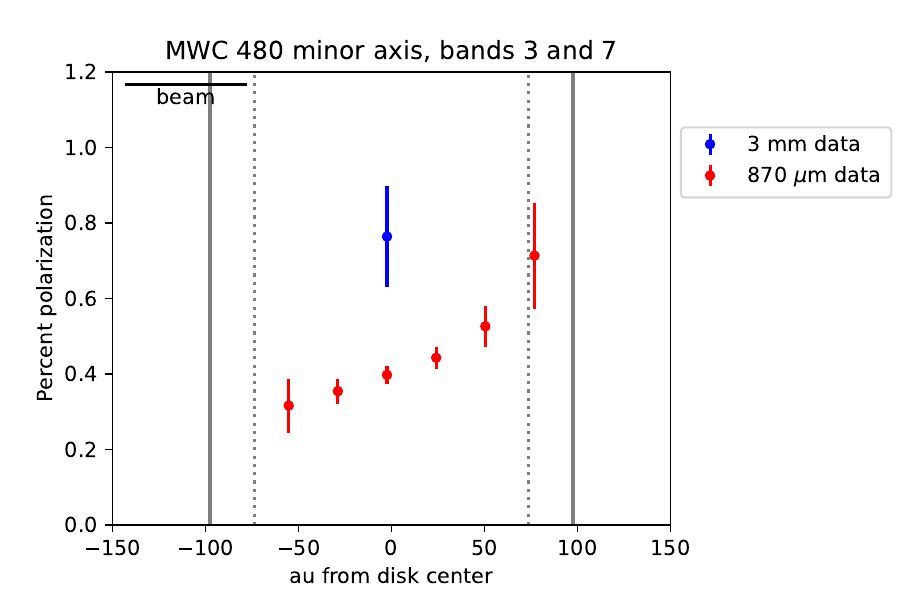}{0.48\textwidth}{(b)}}
\gridline{\fig{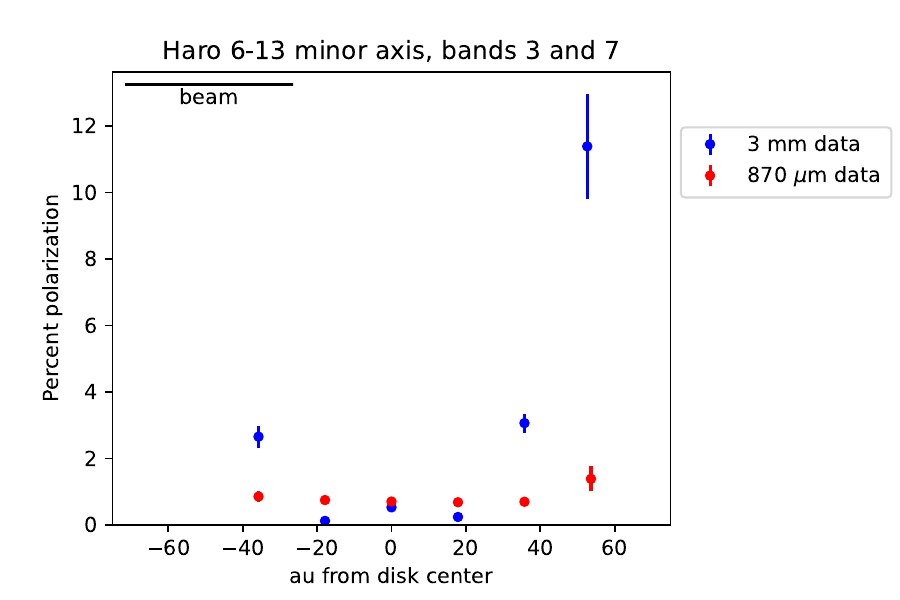}{0.48\textwidth}{(c)}
          \fig{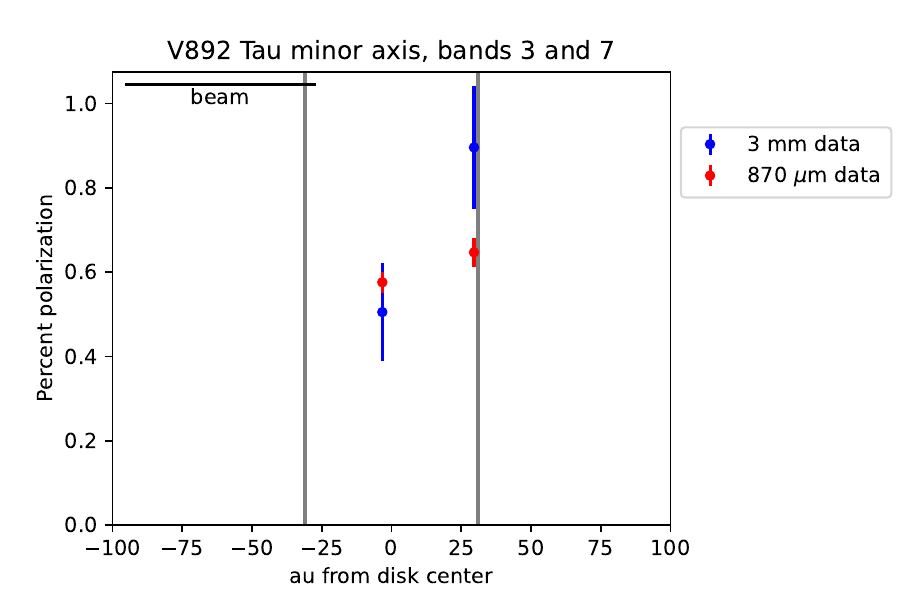}{0.48\textwidth}{(d)}}
\gridline{\fig{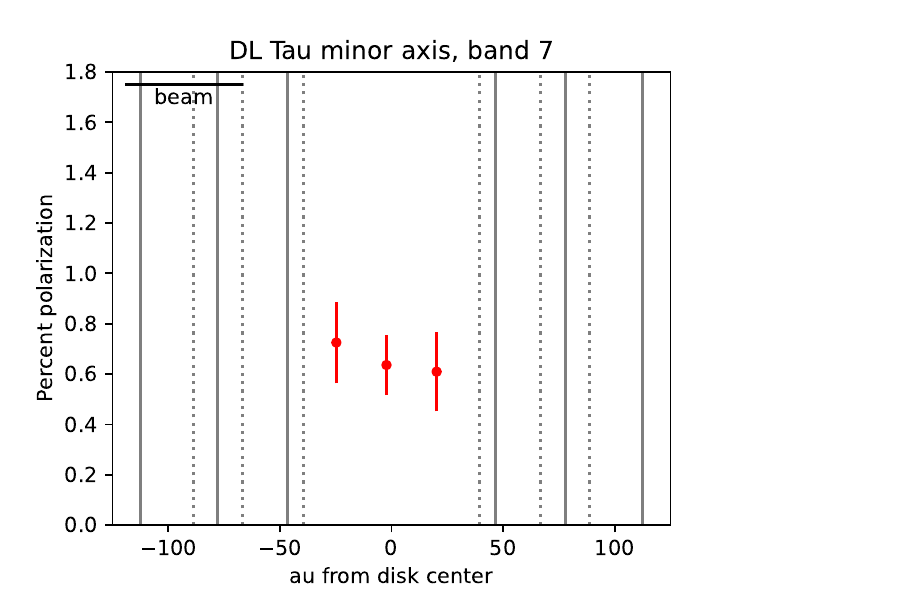}{0.48\textwidth}{(e)}
          \fig{DL_Tau_minax_pper_band3_upper_lim_new.pdf}{0.48\textwidth}{(d)}}

\caption{Percent polarization vs. distance from disk center along the disk major axis at 3 mm and 870 $\mu$m for RY Tau, MWC 480, Haro 6-13, and V892 Tau, and at 870 $\mu$m for DL Tau. For RY Tau, MWC 480, Haro 6-13, and V892 Tau, the two bands have been smoothed to the same resolution; the scale bar indicates the beam size. The smoothed resolutions are listed in Table \ref{tab:3mm_res}. For DL Tau, we show the 870 $\mu$m data at the resolution listed in Table \ref{obs_870}. Since the polarized region in DL Tau at 3 mm was less than half of the beam FWHM across, we have plotted the 3$\sigma$ upper limit on the percent polarization in DL Tau at 3 mm. The resolution at 3 mm is listed in Table \ref{tab:3mm_res}. The data are only plotted in regions where the polarized intensity is at least 3$\sigma_P$. Solid vertical lines indicate the location of known rings, and dotted vertical lines indicate the location of known gaps.}
\label{twoband_pfrac_minax}
\end{figure*}

Figures \ref{twoband_pfrac_majax} and \ref{twoband_pfrac_minax} show the polarization fractions of each disk vs. distance from the Stokes $I$ center along the disks' major and deprojected minor axes, respectively. In Haro 6-13, MWC 480, RY Tau, and DL Tau, the Stokes $I$ center is defined as the location of the Stokes $I$ peak. In V892 Tau, the Stokes $I$ center is defined as the midpoint between the two Stokes $I$ peaks. Each point in the plots represents the average value of a line of pixels the length of one-half of the beam width projected along the major and minor axes. To determine the value of the points, we calculated the average values of $I$, $Q$, and $U$ in each half-beam-width line of pixels along each axis, then calculated the percent polarization $p$ as 

\begin{equation}
    p = \cfrac{100}{I} \sqrt{Q^2 + U^2}
\end{equation}

To determine the uncertainties on $I$, $Q$, and $U$, we calculated the average value of sixteen rectangular regions half the size of the beam in regions of the image outside of any emission. We take the variance of those means as the uncertainty on the average values of $I$, $Q$, and $U$. We calculate the uncertainty on the percent polarization, $\sigma_p$ as 

\begin{align*}
    \sigma_p = \cfrac{100}{I} \Bigg( \Bigg. & 
        \cfrac{1}{(Q^2+U^2)}[(Q\sigma_Q)^2 + (U\sigma_u)^2] \\
    & +\left[\left(\cfrac{Q}{I}\right)^2 + \left(\cfrac{U}{I}\right)^2\right]\sigma_I^2 \Bigg. \Bigg)^{1/2}
\label{eq:sigp}
\end{align*}

The debiased percent polarization, $p'$, which is plotted in Figures \ref{twoband_pfrac_majax} and \ref{twoband_pfrac_minax} is calculated as 

\begin{equation}
    p' = \sqrt{p^2 - \sigma_p^2}
\end{equation} 

In disks known to have rings and gaps, ring locations are indicated by solid vertical lines, and gap locations are indicated by dashed vertical lines. The black scale bar indicates the beam size in the smoothed images.

It is immediately clear from Figure \ref{twoband_composite} that the disks in this survey exhibit two different morphological groups across two wavelengths. 
Of the four disks in which we have observed polarized emission at the 3-$\sigma$ level at 3 mm and 870 $\mu$m, Haro 6-13 and V892 Tau show a transition of polarization patterns similar to HL Tau at the same wavelengths, while MWC 480 and RY Tau show polarization patterns indicative of scattering at both wavelengths (i.e., polarization vectors well-aligned to the disk minor axis, as described in \citet{2016MNRAS.456.2794Y}). Additionally, we observed polarization consistent with scattering at 870 $\mu$m in DL Tau.

One notable feature of RY Tau and MWC 480's polarization spectra is the relative polarization fractions at the two wavelengths. Both wavelengths have polarization fractions $>$0.4\%, with a higher polarization fraction at 3 mm in the center of the disk. This is in contrast to previous scattering models such as those presented in \citet{2017ApJ...844L...5K}, which predict a steep correlation between polarization fraction and observing wavelength. In the scattering polarization models for RY Tau and MWC 480 discussed in Section \ref{sec:discussion}, we have aimed to reproduce an observable degree of scattering polarization at both observing wavelengths, and a higher polarization fraction at 3 mm in the center of the disk. 

\section{Discussion} \label{sec:discussion}
The morphological differences in the polarization spectra of our sources are of interest because they may point toward differences in the disks' dust environments and/or correlations between polarization and protostar properties. Dust grain sizes and optical depths affect the degree of scattering polarization observed, as well whether polarization from scattering or from aligned grains dominates. 
In Section \ref{sec:modeling}, we compare the data from RY Tau and MWC 480 to models of the scattering polarization for various dust grain populations. In Section \ref{sec:spectra}, we explore possible causes of the differences in polarization morphology transitions between the two wavelengths. 

Because scattering efficiency at a given wavelength is heavily dependent on dust grain size \citep{2015ApJ...809...78K}, the polarization fraction from scattering is expected to vary significantly with observing wavelength for a disk whose dust grain sizes are defined by a simple power-law distribution \citep{2017ApJ...844L...5K}. However, MWC 480 and RY Tau show a fairly similar degree of scattering polarization at two widely-separated observing wavelengths. Several scenarios that could give rise to the significant degree of polarization in these sources at 870 $\mu$m and 3 mm, including:
\begin{enumerate}
    \item \textit{Optical depth effects: }The dust is well-described by a single population with a defined maximum grain size. Polarization fraction depends on the optical depth, which differs at different wavelengths. In the optically thin limit, the polarization fraction increases with increasing optical depth. The polarization peaks when the optical depth is of order unity after which decreases to a constant value in the optically thick limit (Yang et al. 2017). If the disk is optically thick at 870 $\mu$m, optical depth effects would attenuate scattering polarization at that wavelength. 
    \item \textit{Multiple dust grain populations: }The disks contain multiple dust grain populations with different maximum sizes. The vertical component of the stellar gravity pulls the grains towards the disk midplane, causing them to settle, and this effect is stronger for larger grains \citep{2005A&A...443..185B}. The larger grains dominate the optical depth at longer wavelengths, whereas the smaller grains are effectively invisible and do not contribute to polarization. At shorter wavelengths, increase in the optical depth of the smaller grains that are elevated can screen out emission of the large grains near the midplane \citep{Ueda2021}, meaning that scattering polarization from smaller grains above the midplane would dominate.
    \item \textit{Dust grain porosity: }The dust grains in the disk are highly porous. Scattering from porous dust grains leads to a flatter polarization spectrum than scattering from non-porous grains (see, e.g., Figure 10 of \citealt{Tazaki2019}). 
    \item \textit{Disk substructure: }Ring/gap structures in the disk create radial variations in optical depth. These variations in optical depth then affect the observed polarization fraction from scattering, as described in \citet{2020MNRAS.496..169L}.
\end{enumerate}

Investigating the effects of disk substructure on these sources' polarization spectra will require higher resolution observations. At the native resolution of the 870 $\mu$m images, only the ring and gap in MWC 480 are resolved. Higher sensitivity observations will also be necessary to explore the effects of disk substructure; the polarized region of MWC 480 only extends to the disk's gap, and higher sensitivity observations would reveal fainter polarized emission at larger radii. Since dust porosity is poorly constrained and disk substructure is not well resolved in our sources, we will explore only scenarios 1 and 2 further by comparing the 3 mm and 870 $\mu$m data to models of the polarized emission produced by scattering for several different dust grain populations in Section \ref{sec:modeling}. 

\subsection{Scattering models}\label{sec:modeling}
Since MWC 480 and RY Tau show evidence of pure scattering at both wavelengths, we explore models of these two sources' polarization in this section. While determining the precise dust grain distribution that best fits the data is beyond the scope of this paper, we aim to show that dust settling and optical depth effects should be taken into account when using scattering polarization observations to constrain a disk's dust properties. The key features we aim to reproduce in the models are (1) a significant degree ($\sim$0.5\% - 2\%) of polarization at observing wavelengths of both 3 mm and 870 $\mu$m and (2) a higher polarization fraction for 3 mm than 870 $\mu$m in the center of the disk. These features are of particular interest because they differ from the predictions of simple scattering models such as those in \citet{2017ApJ...844L...5K}. Given that polarization due to scattering depends on inclination (Yang et al. 2016), we use the two inclination angles for the two sources. 

Modeling the scattering polarization for a range of dust populations will help to determine promising avenues for future modeling to further constrain the dust grain sizes in RY Tau and MWC 480. Our models are not fine-tuned to fit the data; instead, our focus is to reproduce features (1) and (2) from the paragraph above. In fact, the exact dust population cannot be constrained from observations of two wavelengths alone, as scattering polarization depends not only on dust grain sizes but on optical depth, albedo, porosity, shape, and temperature. We model the scattering polarization for dust grain populations with maximum radii of 140, 200, and 490 $\mu$m, as well as for a disk with two vertically stratified dust grain populations with maximum radii of 140 and 490 $\mu$m. The single-population models represent dust populations whose scattering polarization fractions would peak at 870 $\mu$m, 1.3 mm, and 3 mm, respectively, if $a_{max} \sim \lambda/2\pi$, as in \citep{2015ApJ...809...78K}. For the two-population model, we have chosen dust grain sizes we know will contribute to the scattering polarization at both observed wavelengths. The two-population model is designed as a proof of concept to show that size-dependent dust settling could produce scattering polarization two widely separated wavelengths, with a higher polarization fraction at the longer wavelength. 

 For the single-population models, we consider models with a characteristic optical depth ($\tau_0$) of 1 or 10. For each single-grain population model, we also consider two ways of determining dust scale height: fixing the dust scale height at 10\% of the gas scale height, or setting the $\alpha$ viscosity parameter to a fixed level of $10^{-3}$ and solve for the dust scale height (Eq.~(\ref{eq:dust_settle_height})). The $\alpha$ parameter is usually estimated to be $\sim 10^{-3} - 10^{-2}$ from magneto-rotational instability \citep[e.g.][]{Simon2015}, but it can be also be as low as $\sim 10^{-4}$ if the gas is weakly coupled to the magnetic field and hydrodynamical effects like the vertical shear instability dominate \citep[e.g][]{Flock2020}. Values inferred from observations also produce a wide range from $\sim 10^{-5}$ \citep{Villenave2022} to $\sim 10^{-4} - 10^{-3}$ \citep[e.g.][]{Boneberg2016, Pinte2016, Ohashi2019, Flaherty2020} and even be as high as $\sim 0.08$ \citep{Flaherty2020}. We pick $10^{-3}$ as a representative value. A viscosity parameter of $10^{-3}$ yields very similar results to setting $H_{dust} = 0.1H_{gas}$. Decreasing $\alpha$ to $10^{-5}$ decreases the $H_{dust}/H_{gas}$ by a factor of 10, but \citep{Ueda2021} find that size-dependent vertical dust settling occurs for $\alpha > 10^{-5}$. 

In the two-population model, each dust population is treated as having its own characteristic optical depth; the optical depth of a population is what the optical depth of the disk would be if only that dust grain population were present. The $\alpha$ viscosity parameter in the two-population model is set to $10^{-3}$, and we then solve for the scale height of each dust population. 
We consider a two-population model in which the characteristic optical depths of the 140 $\mu$m and 490 $\mu$m populations are 9 and 1 respectively. 

We use the radiative transfer code RADMC-3D \citep{radmc} to calculate the polarization. Our model assumes compact spherical grains, and we adopt the dust mixture from the Disk Substructures at High Angular Resolution Project (DSHARP) prescribed in Birnstiel et al. (2018) \footnote{The code for calculating the opacity is available at \url{https://github.com/birnstiel/dsharp_opac}.}. The composition of the dust (and its refractive index) is water ice (\citealt{2008JGRD..11314220W}), astronomical silicates (\citealt{2003ARA&A..41..241D}), troilite (\citealt{1996A&A...311..291H}), and refractory organic material (\citealt{1996A&A...311..291H}) with mass fractions of $\sim 0.2$, $0.33$, $0.07$, and $0.4$, respectively. For a dust population, the grain size $a$ follows a power-law distribution which goes as $a^{-3.5}$ \citep{1977ApJ...217..425M} with cut-offs at a minimum grain size of 0.1~$\mu$m and a maximum grain size $\amax$ which we treat as a free parameter. Averaging each dust population over a size distribution as opposed to using a single grain size helps produce a smooth scattering matrix and avoids quick oscillations as a function of the scattering angle. 

For the disk structure, we parameterize the radial temperature profile as 
\begin{align}
    T(r) = 25 \left(\cfrac{r}{100 \text{au}}\right)^{-0.5} \text{K}
\end{align}
which is motivated by results from irradiated flared disks in radiative equilibrium (e.g., \citealt{1997ApJ...490..368C}). 
The surface density of the $i$th dust population is parameterized by a simple prescription \citep{1974MNRAS.168..603L}:
\begin{align} \label{eq:dust_surface_density}
    \Sigma_{i}(r) = \dfrac{ \mu \tau_{0,i} }{ \kappa_{\text{ext}, i} } \left(\cfrac{r}{50 au}\right)^{-0.5} \exp \bigg[ - \bigg(\dfrac{r}{ 50 \text{au} }\bigg)^{1.5} \bigg],
\end{align}
where $\mu \equiv \cos \theta$. $\kappa_{\text{ext}, i}$ is the extinction opacity and is calculated from the dust model which for our case, only depends on $\amax$. $\tau_{0,i}$ is the characteristic optical depth at Band~7 for a given dust population. It is the optical depth at Band~7 (at 50~au modulated by a factor of $\sim 0.37$ from the exponential taper as shown in Eq.~(\ref{eq:dust_surface_density})) if only that one dust population were present and viewed face-on. We treat $\tau_{0,i}$ as a free parameter for our exploration below.

For simplicity, the vertical density distribution of the $i$th grain population follows a Gaussian
\begin{align}
    \rho_{i}(r,z) = \dfrac{ \Sigma_{i} }{ \sqrt{2 \pi} H_{i} } \exp \bigg[ - \dfrac{1}{2} \bigg( \dfrac{z}{H_{i}} \bigg)^{2} \bigg],
\end{align}
where $H_{i}$ is the dust scale height. We consider dust settling by assuming that dust is in equilibrium between turbulent diffusion and gravitational settling. As such, the dust scale height depends on the size through 
\begin{align} \label{eq:dust_settle_height}
    \dfrac{ H_{i} }{ H_{g} } = \bigg( 1 + \dfrac{\text{St}}{\alpha} \dfrac{1 + 2 \text{St}}{1 + \text{St}} \bigg)^{-1/2},
\end{align}
\citep[e.g.][]{1995Icar..114..237D, 2007Icar..192..588Y, 2019ApJ...886..103O}
where $\text{St}$ is the Stokes number and $\alpha$ is the dimensionless parameter that describes the level of turbulence \citep{1973A&A....24..337S}. The gas scale height is simply $H_{g} = c_{s} / \Omega$ where $c_{s}$ is the isothermal sound speed and $\Omega \equiv \sqrt{GM / r^{3}}$ is the Keplerian frequency. We adopt $M= 2 M_{\odot}$ which is representative of MWC 480 and RY Tau \citep{2018ApJ...869...17L}. 
The Stokes number of the grains in a disk depends on the surface density of the gas through \citep[e.g.][]{2018ApJ...869L..46D}: 
\begin{align}
    \text{St} = \dfrac{\pi}{2} \dfrac{\rho_{s} a }{\Sigma_{g}},
\end{align}
where $\rho_{s}$ is the specific weight of the dust mixture (which is $\sim 1.68$ g cm$^{-3}$ for the DSHARP composition) and $a$ is the grain size which we use $\amax$ for the corresponding dust population. The gas surface density $\Sigma_{g}$ is the summation of each dust surface density $\Sigma_{i}$ times the gas-to-dust mass ratio of 100.

Model images used for RY Tau has an inclination of $\theta=65^{\circ}$ and those for MWC 480 has $\theta=37^{\circ}$. To account for finite resolution, each model image is convolved with a $0\farcs3$ circular beam.



\begin{figure*}[ht]
\gridline{\fig{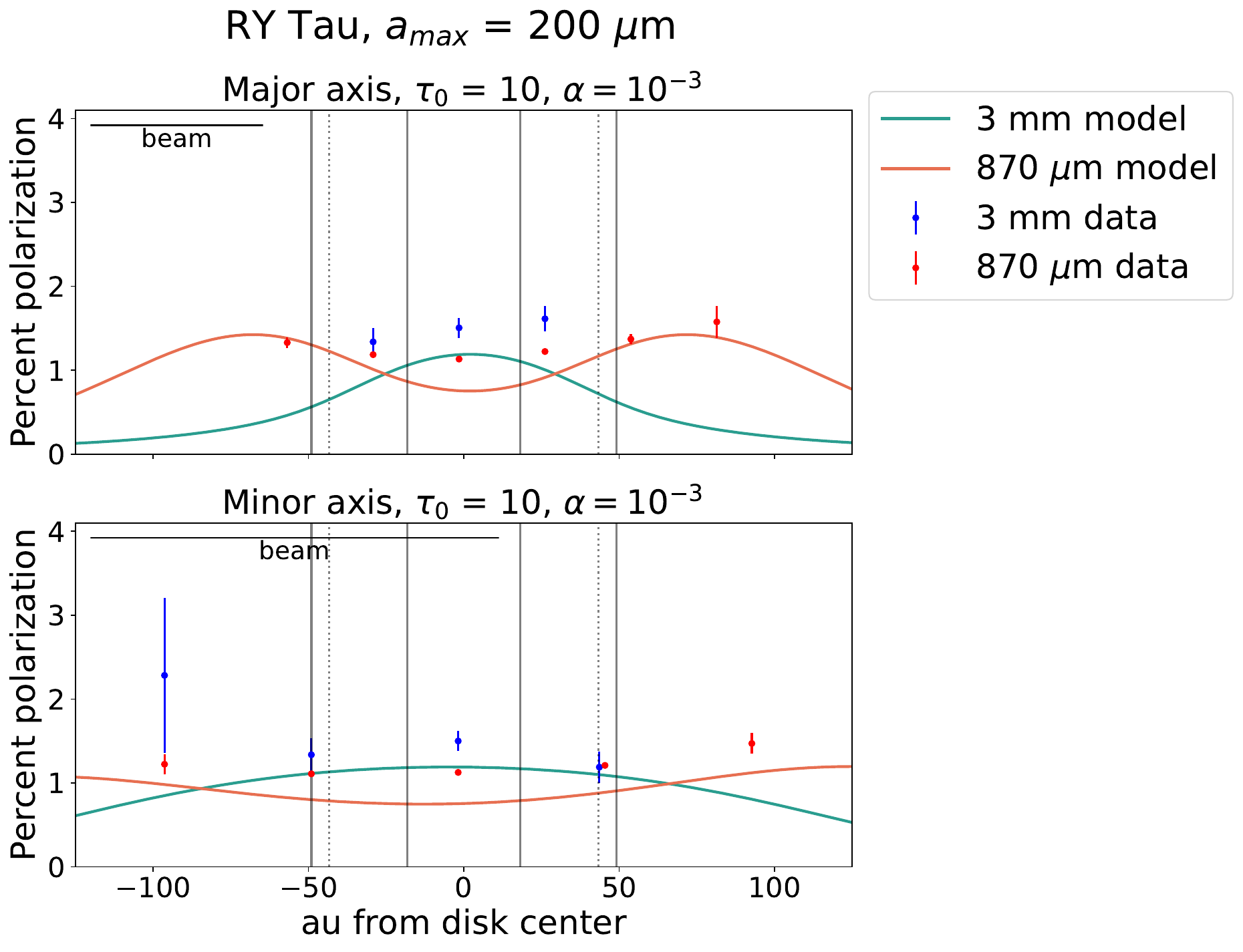}{0.45\textwidth}{(a)}
          \fig{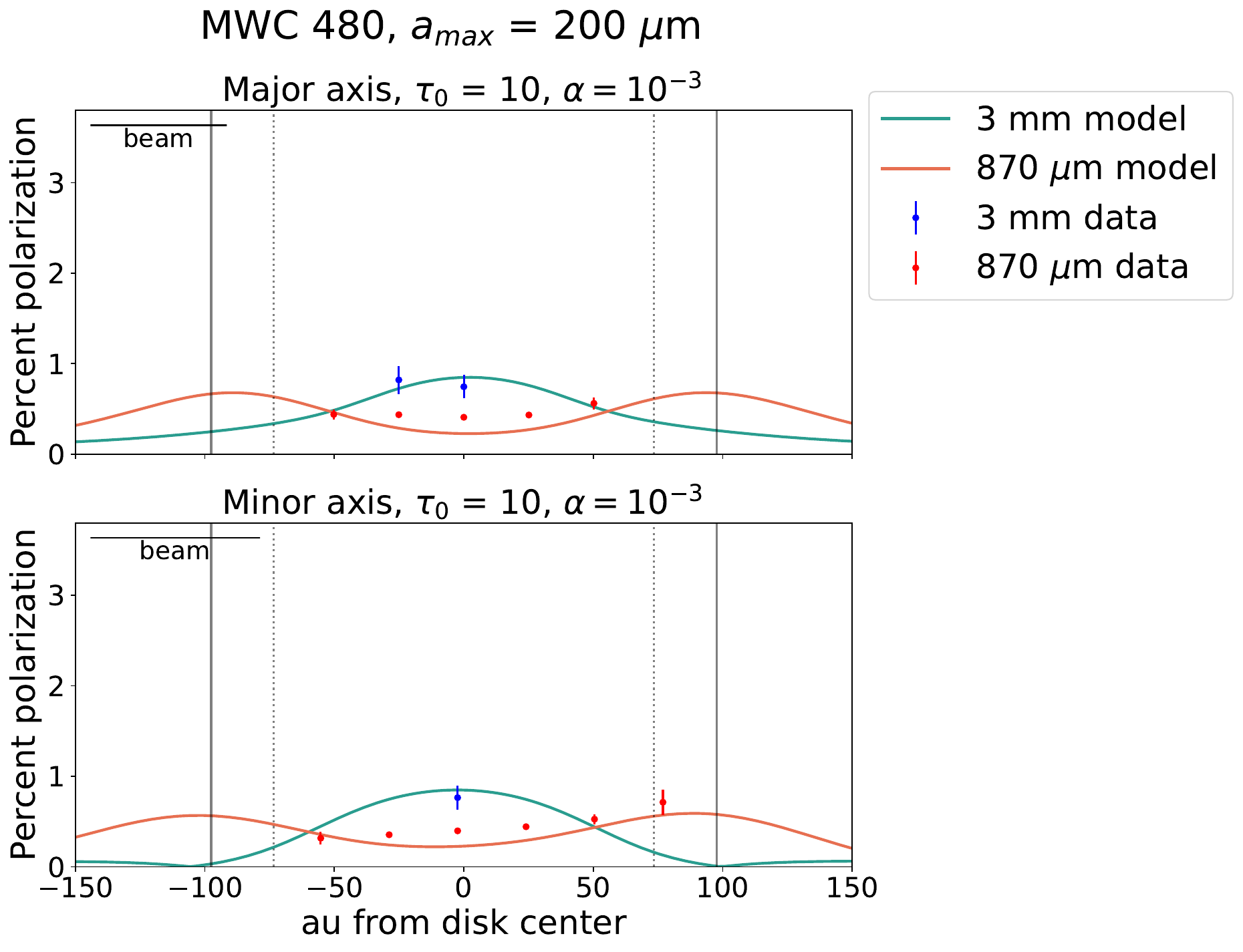}{0.45\textwidth}{(b)}}

\caption{Data vs. scattering model with a maximum grain size of 200 $\mu$m along the major and minor axes of RY Tau and MWC 480. Models have been convolved with a 0$\farcs$3 beam.}
\label{200um_modelvdata}
\end{figure*}

\begin{figure*}[ht]
\gridline{\fig{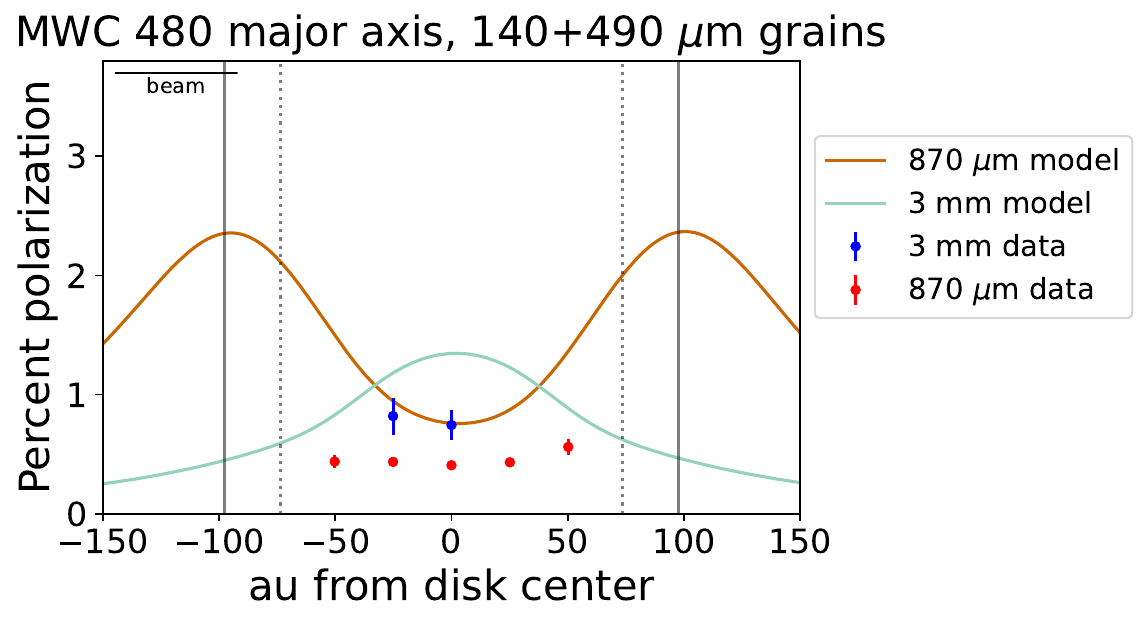}{0.48\textwidth}{(a)}
          \fig{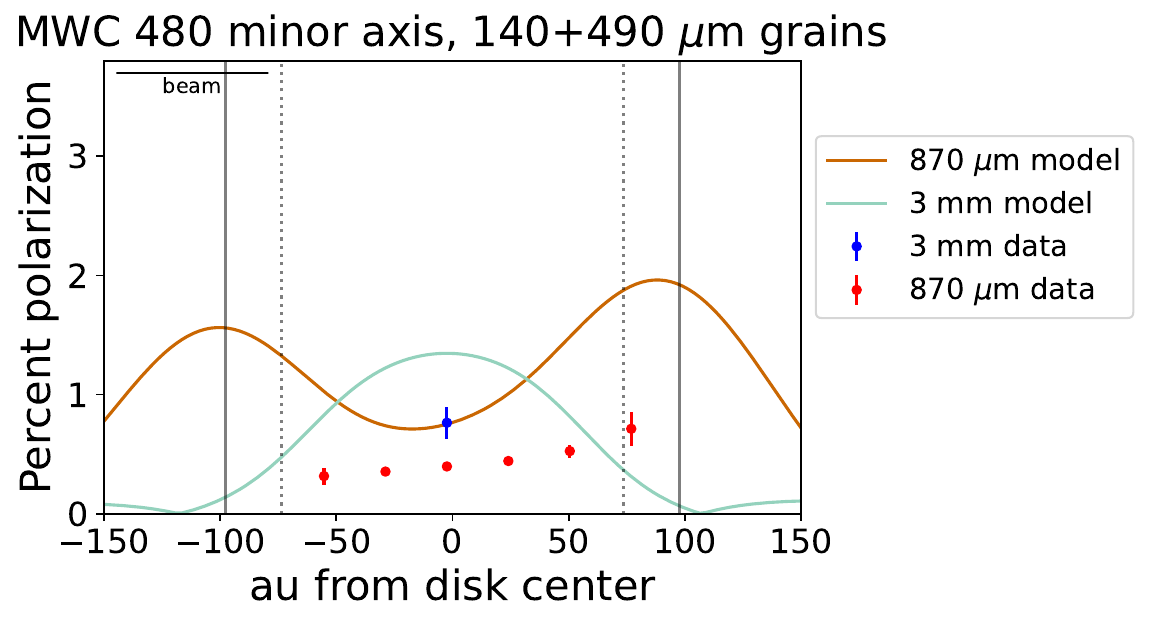}{0.48\textwidth}{(b)}}

\caption{Data vs. scattering model with a maximum grain sizes of 140 and 490 $\mu$m along the major and minor axes of MWC 480. Models have been convolved with a 0$\farcs$3 beam.}
\label{multigrain_modelvdata}
\end{figure*}

Figures \ref{200um_modelvdata} and \ref{multigrain_modelvdata} show the models which produce a detectable degree of polarization at both wavelengths and a higher polarization fraction at 3 mm at the center of the disk. In these figures, as well as the figures in the Appendix, distances along the minor axis have been deprojected to account for the disks' inclination angles. The radial locations of features seen in the models are different along the major and minor axis because the larger deprojected beam FWHM along the minor axis smears out features. In both MWC 480 and RY Tau, the single-population scattering model with a maximum dust grain size of 200 $\mu$m reproduces the key features. The two-population model also reproduces the key features in MWC 480. We note that the effects of dust grain size and optical depth on scattering polarization are degenerate to some extent; high optical depth can decrease the observed polarization fraction from the theoretical maximum value, but so can a maximum dust grain size smaller or larger than the size that would produce the highest polarization fraction at a given observing wavelength. For example, in MWC 480, the two-population model overpredicts the polarization fraction at both wavelengths. Because the effects of optical depth and dust grain size on scattering polarization are somewhat degenerate when data are only available at two wavelengths, adjusting either the optical depths or the maximum dust grain sizes could bring the model results closer to the observed polarization fractions. Future polarization observations at wavelengths between 870 $\mu$m and 3 mm, as well as higher-resolution, higher sensitivity observations, would allow us to fine-tune our models. 

Although our models fit the general trends, they could be brought into closer detailed agreement with the observations by changing the maximum dust grain size(s) to values between 140 and 200 $\mu$m or between 200 and 490 $\mu$m. Increasing the total optical depth to the point where the polarization fraction is attenuated as the radiation field incident on a grain becomes more isotropic could also bring the model polarization fractions in Figures 6 and 7 closer to the observations. Additionally, larger ($\sim$mm-sized) compact dust aggregates could produce an observable degree of scattering polarization at both 3 mm and 870 $\mu$m, as presented in \citet{Tazaki2019}. Future efforts including comparing models of the Stokes I for these dust populations to the observed Stokes I of these disks individually would be valuable for constraining the dust properties of each source.

In the Appendix, we compare the data to the full suite of models. As seen in Figures \ref{singlegrain_rytau_tau1} and \ref{singlegrain_mwc_tau1}, models in which the maximum dust grain size is 140 $\mu$m or 490 $\mu$m produce a significant degree of polarization at 870 $\mu$m and 3 mm, respectively, but not at the other wavelength. In models in which the dust grains have a maximum radius of 200 $\mu$m, both wavelengths show polarization fractions that would be detectable with ALMA. However, only the model with a higher characteristic optical depth ($\tau_0 = 10$) shows a higher polarization fraction at 3 mm than at 870 $\mu$m at the center of the disk (see Figures \ref{singlegrain_rytau_tau10} and \ref{singlegrain_mwc_tau10}). The dip in polarization fraction in the center of the disk at shorter wavelengths is due to optical depth effects. As the disk becomes more optically thick, the incident radiation field becomes more isotropic, and the net polarization fraction from scattering decreases \citep{2017MNRAS.472..373Y}. In the two-population model, a degree of polarization that would be observable with ALMA is produced at both wavelengths in both disks. In MWC 480, the two-population model produces a higher polarization fraction at 3 mm in the center of the disk, while the polarization fraction remains higher at 870 $\mu$m in the center of the disk in RY Tau (see Figure \ref{twograin_model}).

One possible origin for two size populations at a given radius is that one population could be originally associated with the local gas while the other could be dust that has drifted radially (relative to the gas) to that radius. Numerical simulations with detailed dust physics (including growth, fragmentation, radial drift and vertical settling) are needed to determine whether a disk could host a bimodal dust population. The assumed size distribution of the dust grains would affect the expected polarization fractions. For example, \citet{Ueda2021} modeled the scattering polarization in HL Tau for MRN distributions of dust grain sizes with different maximum sizes and different levels of turbulence. They found that, if the dust grain sizes have an MRN distribution with a single maximum size, then grains with a maximum radius of size of $\lesssim$1 mm and turbulence strength parameter of $\lesssim 10^{-5}$ could explain the polarization seen at 870 $\mu$m and 1.3 mm, as well as the disk’s SED.




\subsection{Polarization spectra}
\label{sec:spectra}

The disks in this survey exhibit clear differences in the correlation between polarization morphology and observing wavelength. This may indicate differences in the disks' dust grain sizes or dust optical depths, or it could arise from differences in protostar luminosities. In this section, we examine each of these cases. 

\citet{2022MNRAS.512.3922L} found that the transition between the azimuthal polarization morphology seen in sources such as HL Tau, DG Tau, and Haro 6-13 at 3 mm (which likely arises from dichroic emission of aligned grains) and the scattering polarization seen at 870 $\mu$m in the same source could be explained by optical depth effects. Optical depth influences which polarization mechanism dominates the polarized emission produced at a given wavelength. At low optical depths, the probability that a photon will encounter a dust grain and scatter to produce polarization are low. Therefore, the scattering polarization tends to be weaker than any polarization from direct thermal emission from aligned dust grains. At high optical depths, photons are likely to encounter an aligned dust grain, which preferentially absorb light whose E-vector is along their long axes. This preferential absorption, known as dichroic extinction, reduces the amount of polarized emission from thermal emission from aligned grains. On the other hand, high optical depths mean that a photon is likely to encounter a dust grain and scatter, thus raising the amount of scattering polarization. The scattering polarization at high optical depths is the net result of multiple scattering events. 

While the transition between azimuthal and scattering polarization occurs between 3 mm and 870 $\mu$m in Haro 6-13 and V892 Tau, RY Tau and MWC 480 are still both consistent with scattering at 3 mm. Given the relationship between polarization morphology and optical depth described in \citet{2022MNRAS.512.3922L}, we expect that if there are aligned grains in RY Tau and MWC 480, the transition between thermal and scattering polarization occurs at wavelengths longer than 3 mm. Compared to Haro 6-13, RY Tau and MWC 480 also have brighter disks, which are likely to have  higher optical depths and thus a stronger scattering at 3 mm. V892 Tau's disk is brighter still, but its dust polarization may be affected by a large hole surrounding the binary system. It is also interesting to note that RY Tau and MWC 480 have stellar masses significantly higher than that of Haro 6-13 ($\sim$2$M_\odot$ vs $\sim$1$M_\odot$). 

Another possible reason for the differences in polarization morphology between the disks is that Haro 6-13 and V892 Tau do not contain dust grains large enough to efficiently create scattering polarization at 3 mm, or that the population of large dust grains in these sources is optically thin. We speculate that RY Tau and MWC 480 may have higher optical depths than Haro 6-13 due to their polarization spectra. If the transition between scattering and the azimuthal polarization pattern is controlled by optical depth, as explored in \citet{2022MNRAS.512.3922L}, then disks with higher optical depths would transition from showing polarization consistent with scattering to showing azimuthal polarization at longer wavelengths. Observing these disks in polarization at wavelengths longer than 3 mm would allow us to determine where the change in polarization morphology occurs in RY Tau and MWC 480. 

As described in \citet{2022FrASS...9.3927T}, the radiative alignment torque acting on a grain is directly proportionate to the energy density of the radiaton field.  A more luminous protostar would therefore create a stronger RAT, leading to a higher probability of radiatively-aligned grains. V892 Tau is the most luminous source in our sample (see Table \ref{tab:disk_params}), and shows a transition between an azimuthal polarization morphology at 3 mm and a morphology consistent with scattering at 870 $\mu$m. RY Tau and MWC 480 are more luminous than Haro 6-13, but are consistent with scattering at both wavelengths, while Haro 6-13 shows the same transition of polarization morphologies as V892 Tau. All else being equal, we would expect higher protostar luminosities to lead to more polarization from radiative alignment, but other differences between the systems may have more influence on their polarization morphologies. DL Tau's disk is less bright than RY Tau and MWC 480's, and it's protostar's luminosity is 0.65$L_\odot$. DL Tau's polarization morphology cannot be conclusively determined from the tentative detection of polarization at 3 mm, since the polarized region was smaller than a beam, allowing for only one measurement of the polarization fraction and angle. If the polarized emission at 3 mm is real, then DL Tau's polarization spectrum resembles RY Tau and MWC 480's. However, higher sensitivity observations are needed to determine whether DL Tau has more extended polarized emission at 3 mm. Determining whether stellar mass and luminosity are correlated with polarization spectrum will require multi-wavelength polarization observations of a wider sample of disks.

\section{Conclusions} \label{sec:conclusions}

In this multi-wavelength survey of dust polarization in protoplanetary disks, we have shown that Class II disks in the same molecular cloud can exhibit different transitions between polarization morphologies at observing wavelengths of 3 mm and 870 $\mu$m. The transition between azimuthal and scattering polarization in Haro 6-13 and V892 Tau can be explained by how optical depth affects whether a source's observed polarization comes primarily from thermal emission of aligned grains or self-scattering. Polarization observations at wavelengths longer than 3 mm, possibly with the Next Generation Very Large Array, will be needed to locate where the transition between the two polarization morphologies occurs in RY Tau and MWC 480. 

In contrast to Haro 6-13 and V892 Tau, the polarized emission in RY Tau and MWC 480 is consistent with scattering at both wavelengths. Using models of scattering polarization, we aimed to determine what dust grain population(s) could produce a significant polarization fraction at both wavelengths and a polarization fraction that was higher at 3 mm than at 870 $\mu$m in the center of the disk. We found that in RY Tau, dust grains with a maximum radius ($a_{max}$) of 200 $\mu$m and an optical depth at 50 au ($\tau_0$) of 10 could reproduce these features. In MWW 480, the model with $a_{max}$ = 200 $\mu$m and $\tau_0$ = 10 could reproduce the key features. The model with large grains ($a_{max}$ = 490 $\mu$m, $\tau_0$ = 1) near the midplane and small grains ($a_{max}$ = 140 $\mu$m and $\tau_0$ = 9) above and below the midplane can also reproduce the key features in MWC 480. With two observing wavelengths, the effects of dust grain size and optical depth (i.e., dust grain number density) on the observed polarization are still somewhat degenerate. The region of degeneracy shrinks as the source is observed at more wavelengths, providing stricter constraints on dust grain properties. 

\section{Acknowledgments}
This paper makes use of the following ALMA data: ALMA \#2017.1.00470.S. ALMA is a partnership of ESO (representing its member states), NSF (USA) and NINS (Japan), together with NRC (Canada), MOST and ASIAA (Taiwan), and KASI (Republic of Korea), in cooperation with the Republic of Chile. The Joint ALMA Observatory is operated by ESO, AUI/NRAO and NAOJ. The National Radio Astronomy Observatory is a facility of the National Science Foundation operated under cooperative agreement by Associated Universities, Inc. We thank the anonymous reviewer, whose comments helped to strengthen this manuscript. REH and LWL acknowledge support from NSF AST-1910364. LWL acknowledges support from NSF AST-1910364 and NSF AST-2307844. REH also acknowledges support from the ALMA Student Observing Support (SOS). ZYDL acknowledges support from the Jefferson Foundation, NASA 80NSSC18K1095, and also support from the ALMA Student Observing Support (SOS). ZYL is supported in part by NASA 80NSSC20K0533 and NSF AST-1910106. 

\bibliography{citations}{}
\bibliographystyle{aasjournal}

\newpage

\appendix

Here, we compare the 870 $\mu$m and 3 mm polarization data for MWC 480 and RY Tau to the full range of scattering models we created. As stated in Section \ref{sec:modeling}, the goal of this modeling is not to find a dust population that fits all features seen in the data, but to provide a proof-of-concept that it is important to consider optical depth effects and dust settling when using scattering polarization observations to constrain dust grain sizes. The polarization fractions are generally higher for RY Tau than for MWC 480 due to inclination-induced polarization \citep{2016MNRAS.456.2794Y}.  We find that, generally, models including a single dust grain population with a maximum grain size of 140 $\mu$m or 490 $\mu$m are not good fits to the data. 

\begin{figure*}
\gridline{\fig{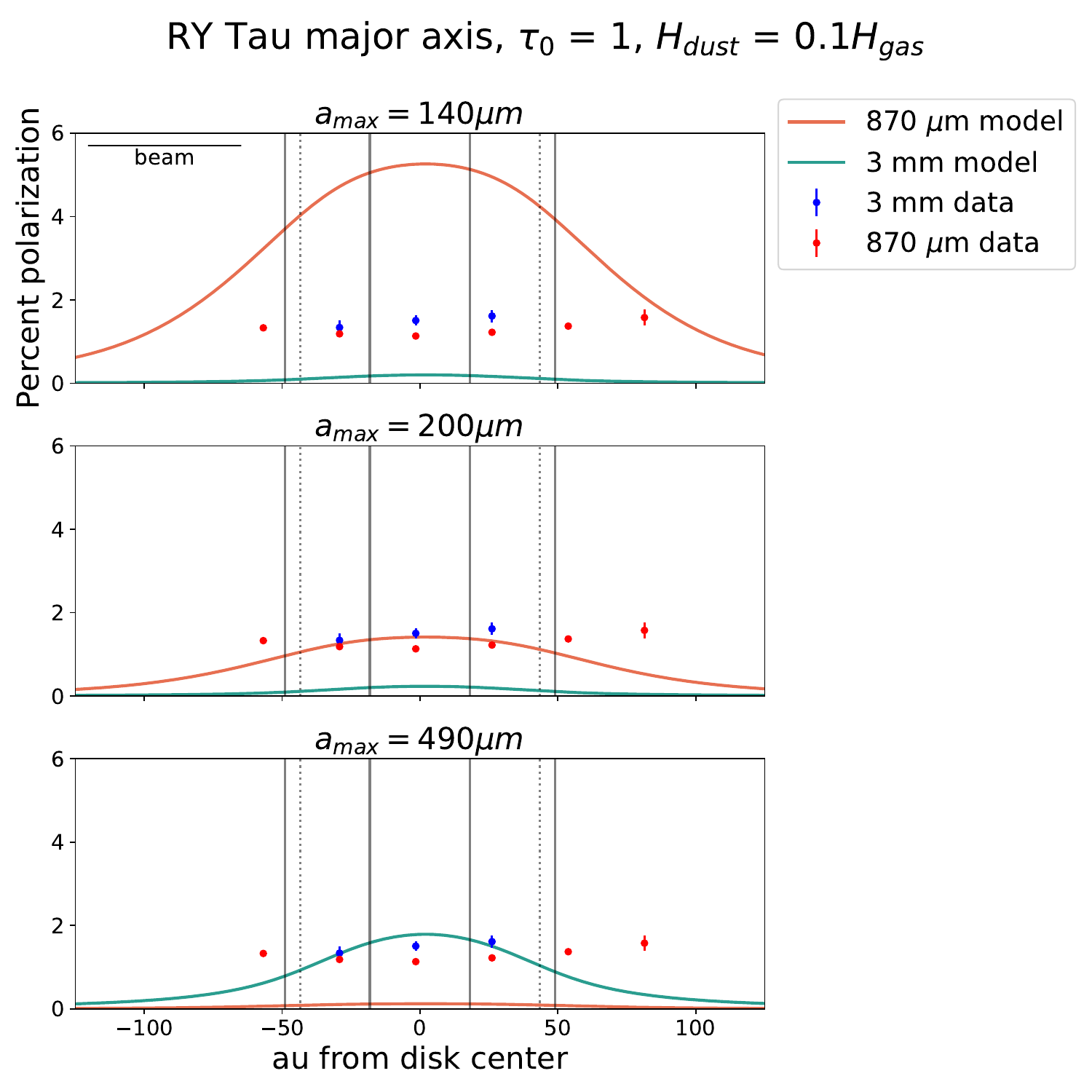}{0.48\textwidth}{(a)}
          \fig{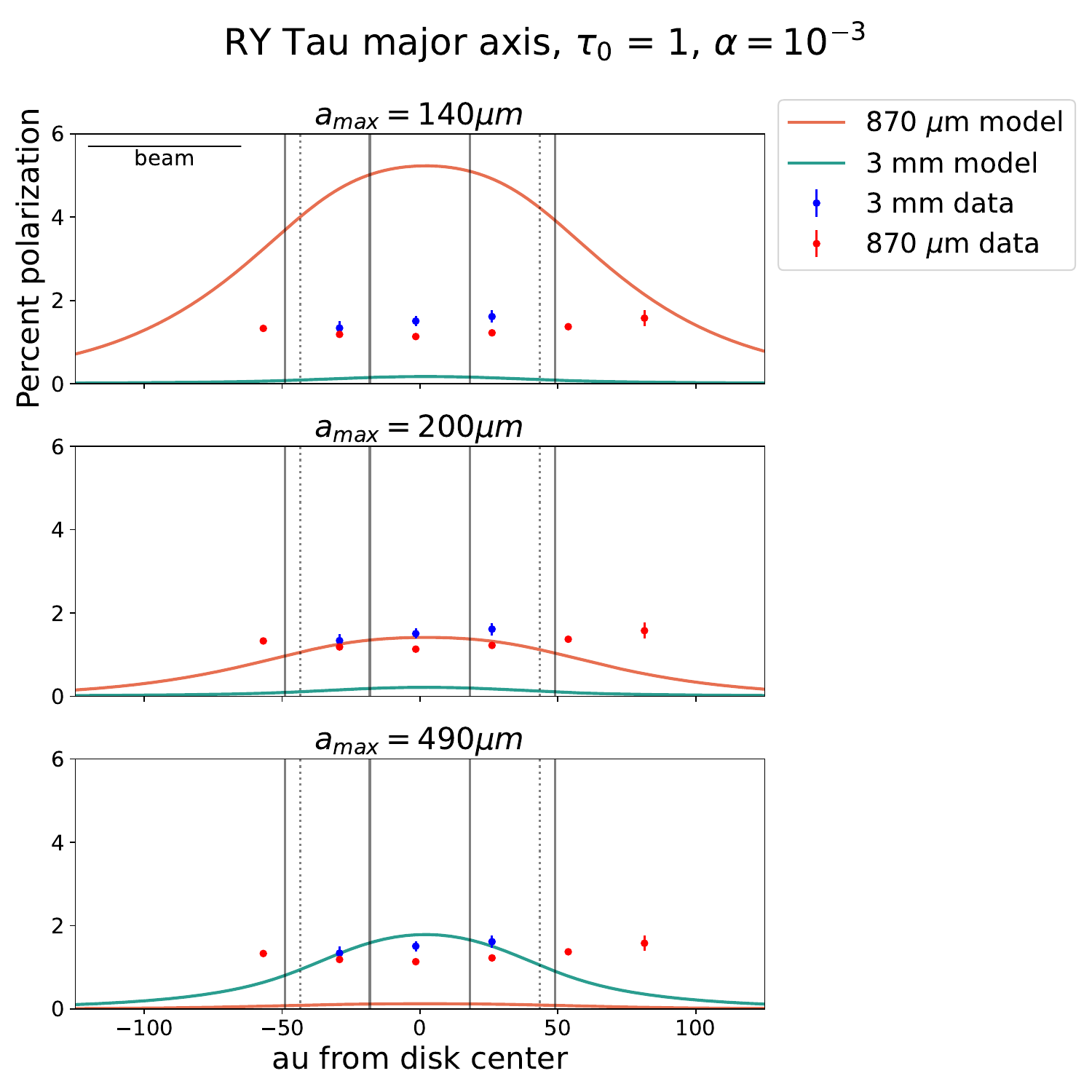}{0.48\textwidth}{(b)}}
\gridline{\fig{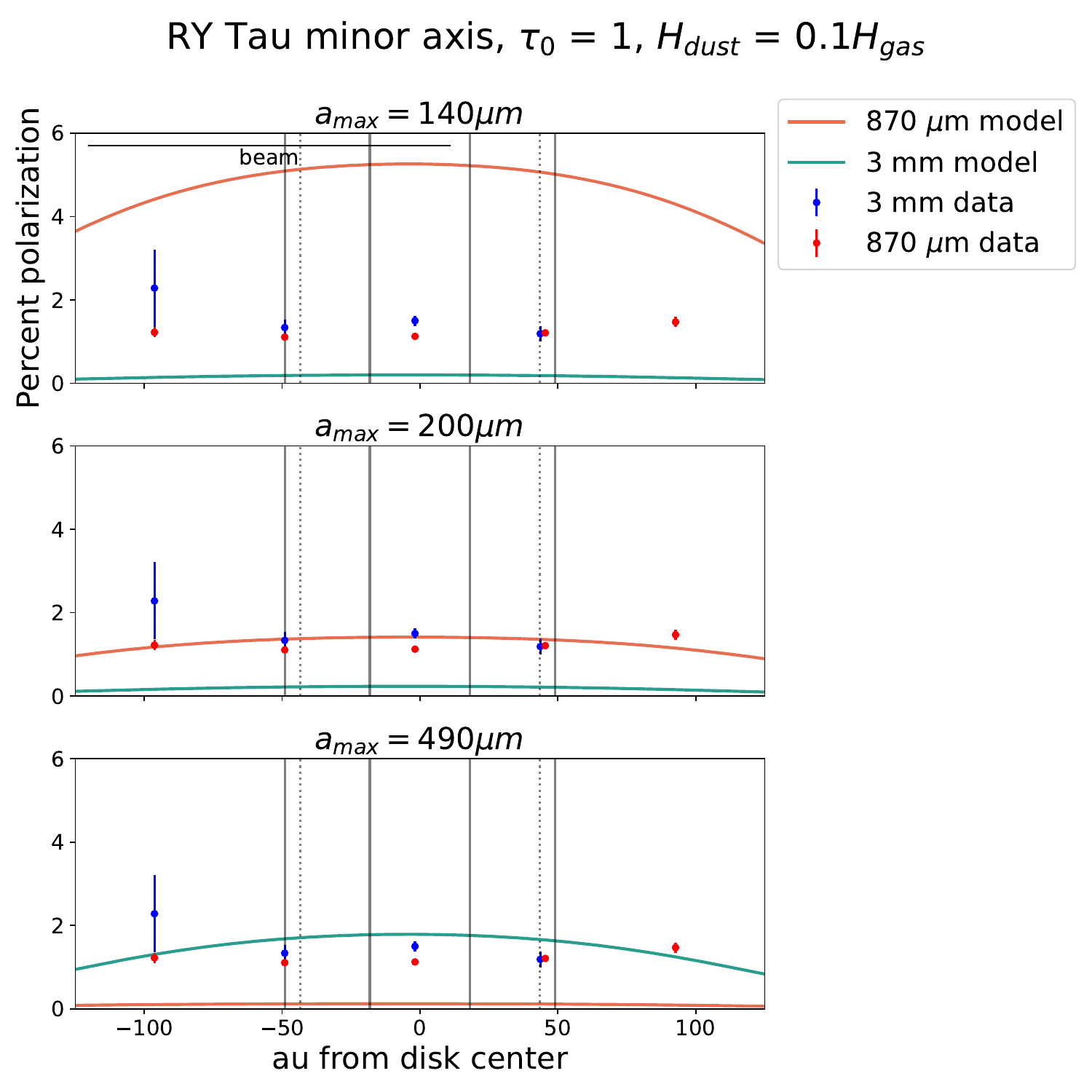}{0.48\textwidth}{(c)}
          \fig{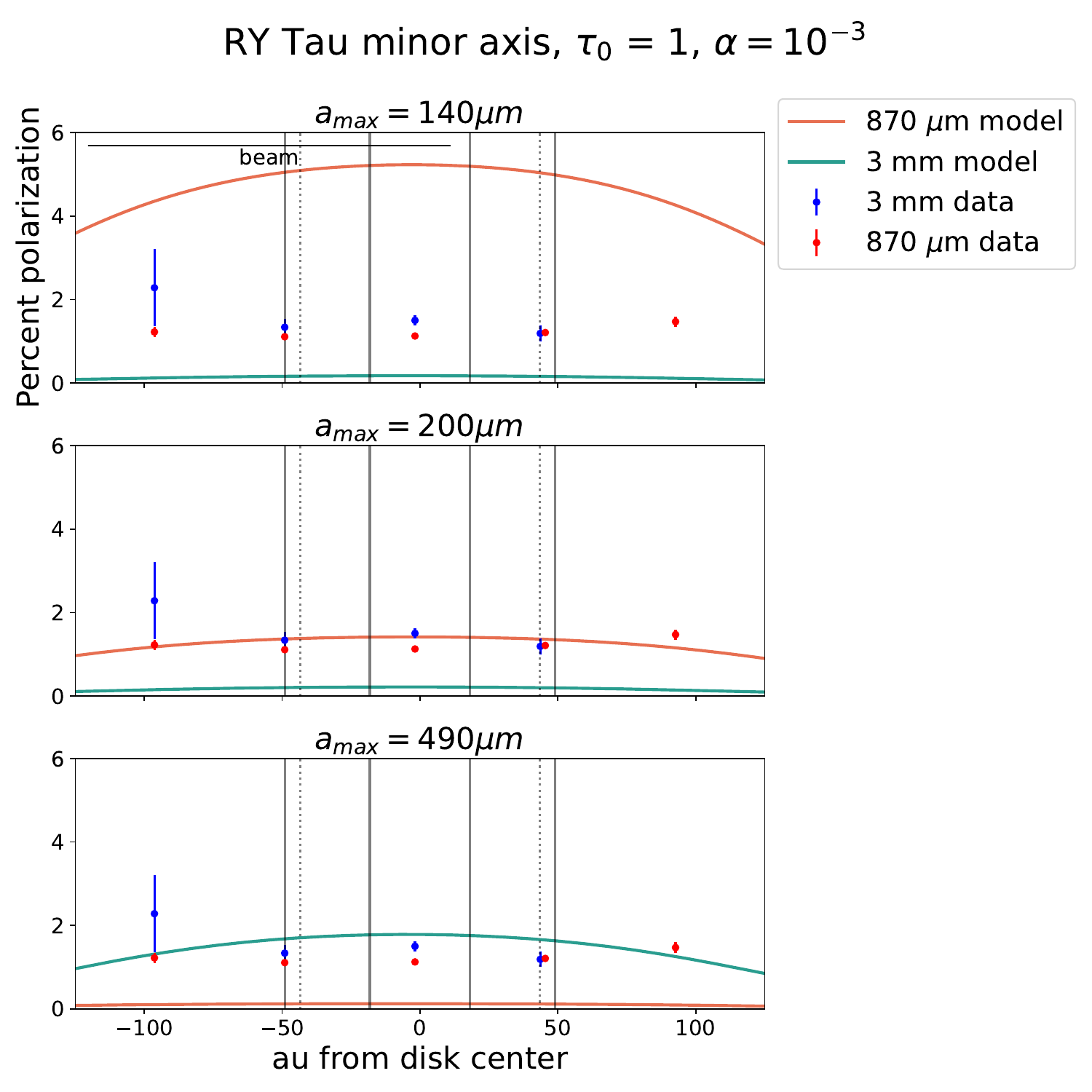}{0.48\textwidth}{(d)}}

\caption{Data vs. $\tau_0=1$ single-grain scattering models along the major and minor axes of RY Tau. Models have been convolved with a 0$\farcs$3 beam. Solid gray lines represent the locations of rings, and dashed gray lines represent the locations of gaps from \citet{2018ApJ...869...17L}. The beam scale bar represents the beam major axis of the 3 mm data.}
\label{singlegrain_rytau_tau1}
\end{figure*}

\begin{figure*}[ht]
\gridline{\fig{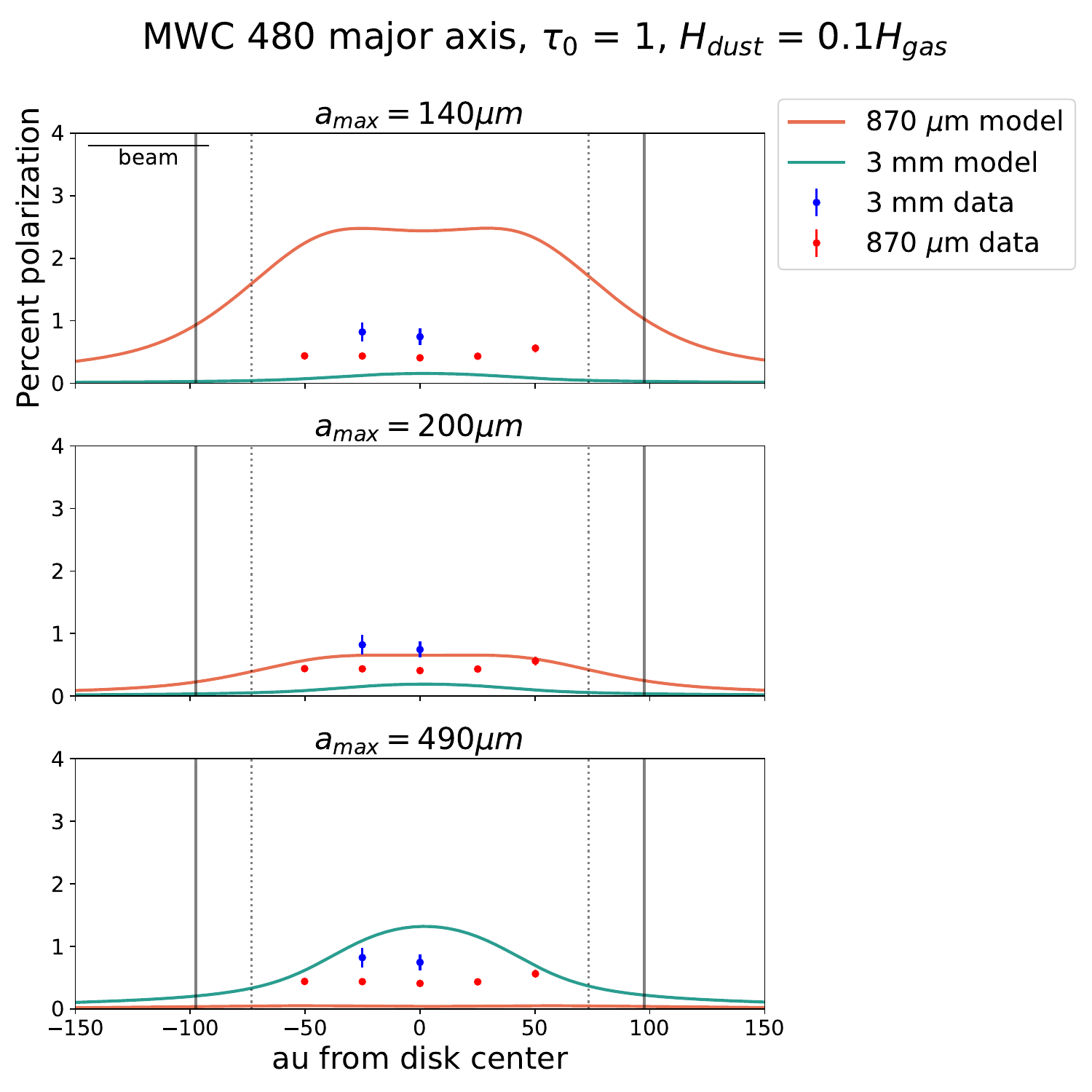}{0.48\textwidth}{(a)}
          \fig{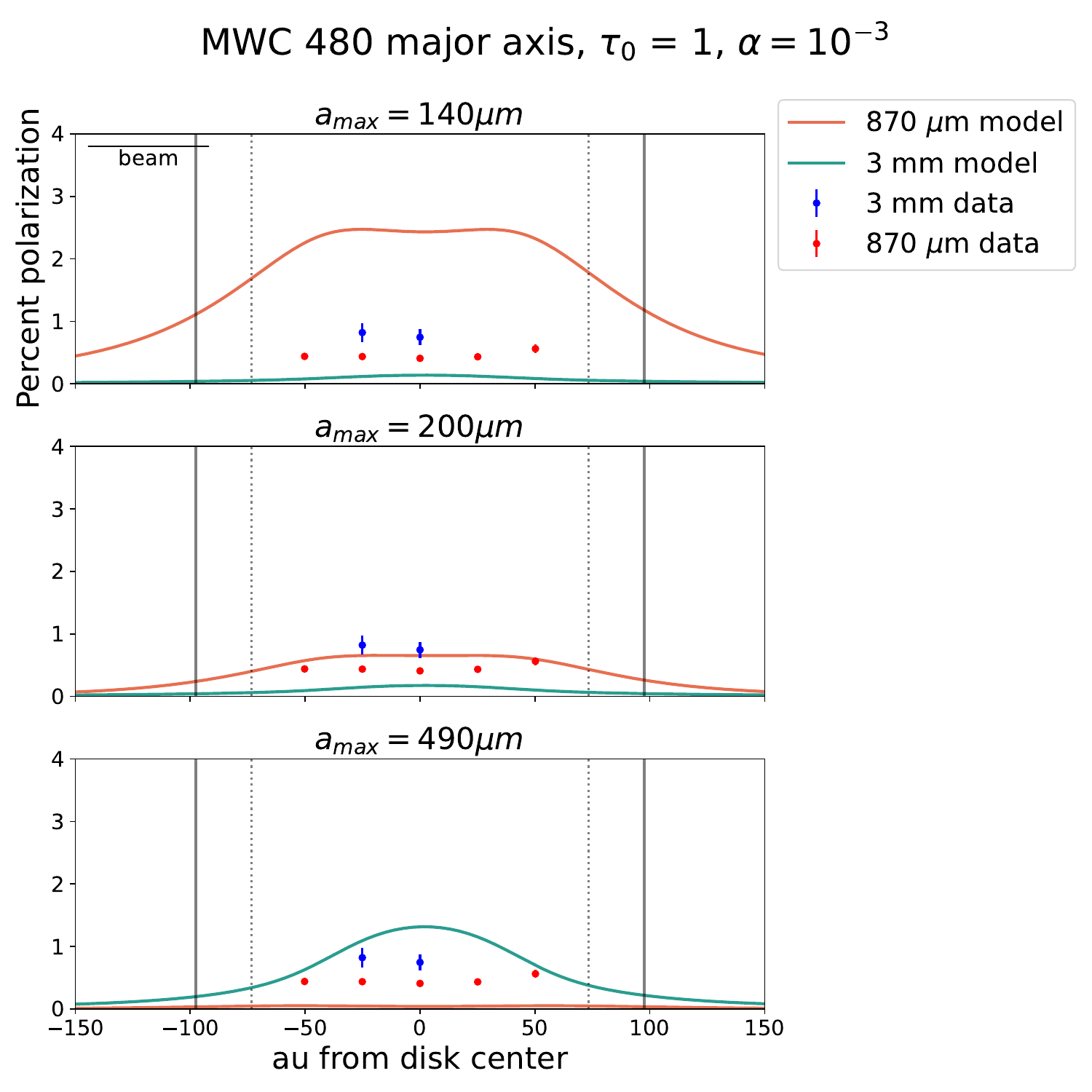}{0.48\textwidth}{(b)}}
\gridline{\fig{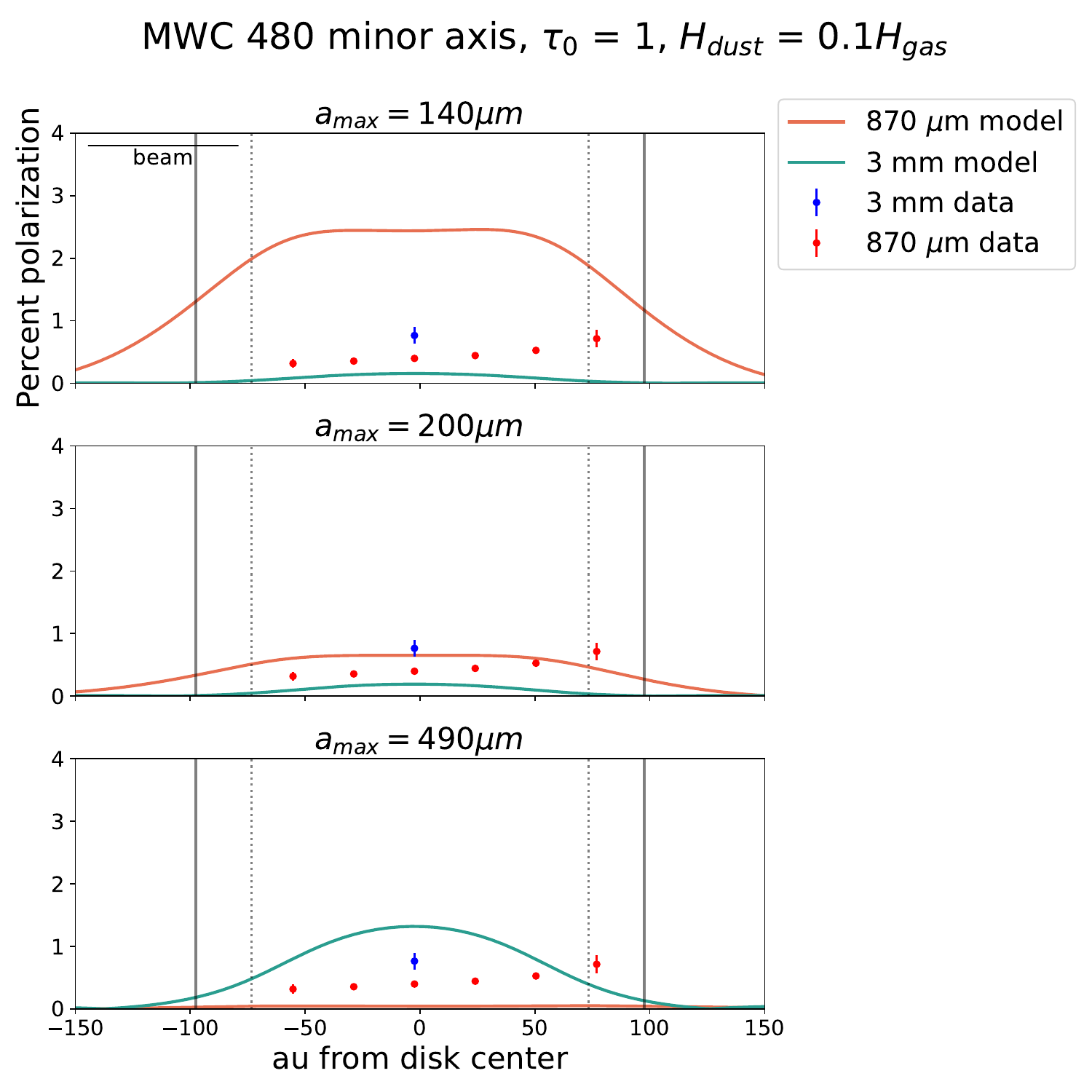}{0.48\textwidth}{(c)}
          \fig{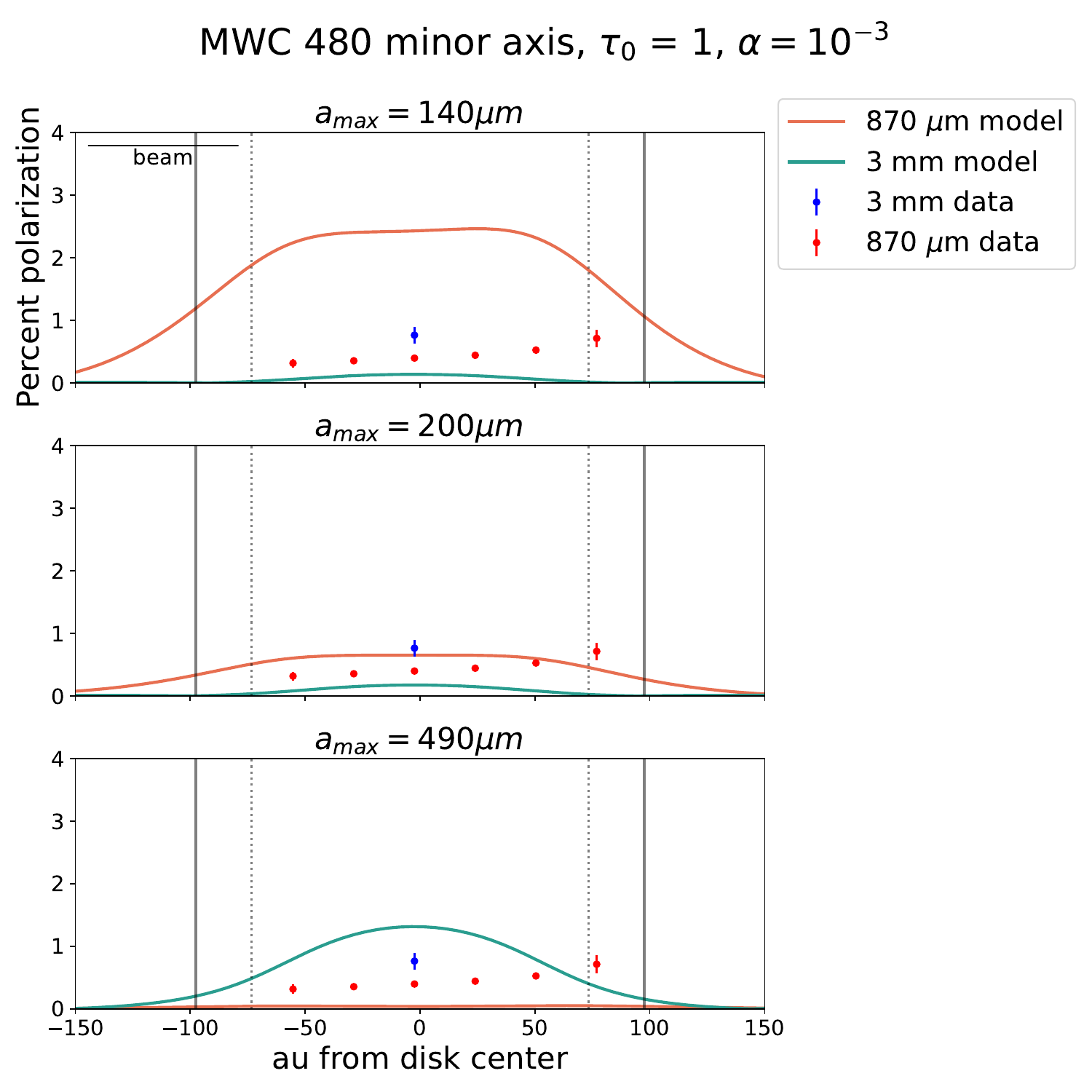}{0.48\textwidth}{(d)}}

\caption{Data vs. $\tau_0=1$ single-grain scattering models along the major and minor axes of MWC 480. Models have been convolved with a 0$\farcs$3 beam. Solid gray lines represent the locations of rings, and dashed gray lines represent the locations of gaps from \citet{2018ApJ...869...17L}. The beam scale bar represents the beam major axis of the 3 mm data.}
\label{singlegrain_mwc_tau1}
\end{figure*}

At low optical depths ($\tau_0=1$), the 140 $\mu$m dust efficiently produces polarized emission at 870 $\mu$m, but produces very low polarization fractions at 3 mm, and vice versa for the 490 $\mu$m dust (see the top and bottom panels, respectively, of Figures \ref{singlegrain_rytau_tau1} and \ref{singlegrain_mwc_tau1}). The polarization fractions produced at 870 $\mu$m and 3 mm by the $a_{max}$ = 200 $\mu$m model with low optical depth are more similar to one another; however, this model still produces a higher polarization fraction at 870 $\mu$m than at 3 mm in the center of the disk, contrary to the data (see the middle panels of Figures \ref{singlegrain_rytau_tau1} and \ref{singlegrain_mwc_tau1}). None of the models with low characteristic optical depth and a single dust grain population reproduce the relative polarization fractions seen in the data. Therefore, we have explored models with high characteristic optical depths and multiple dust grain populations. 

\begin{figure*}
\gridline{\fig{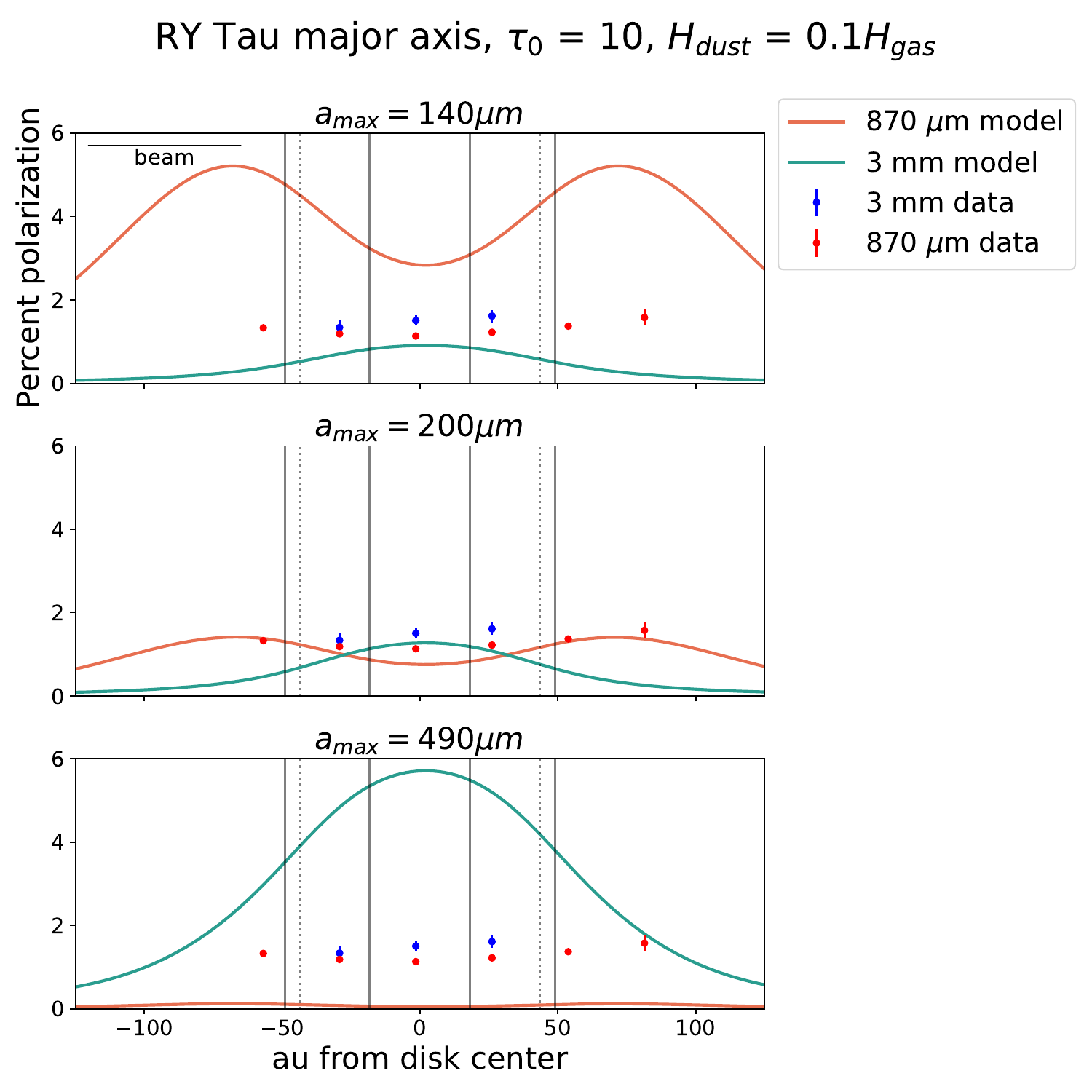}{0.48\textwidth}{(c)}
          \fig{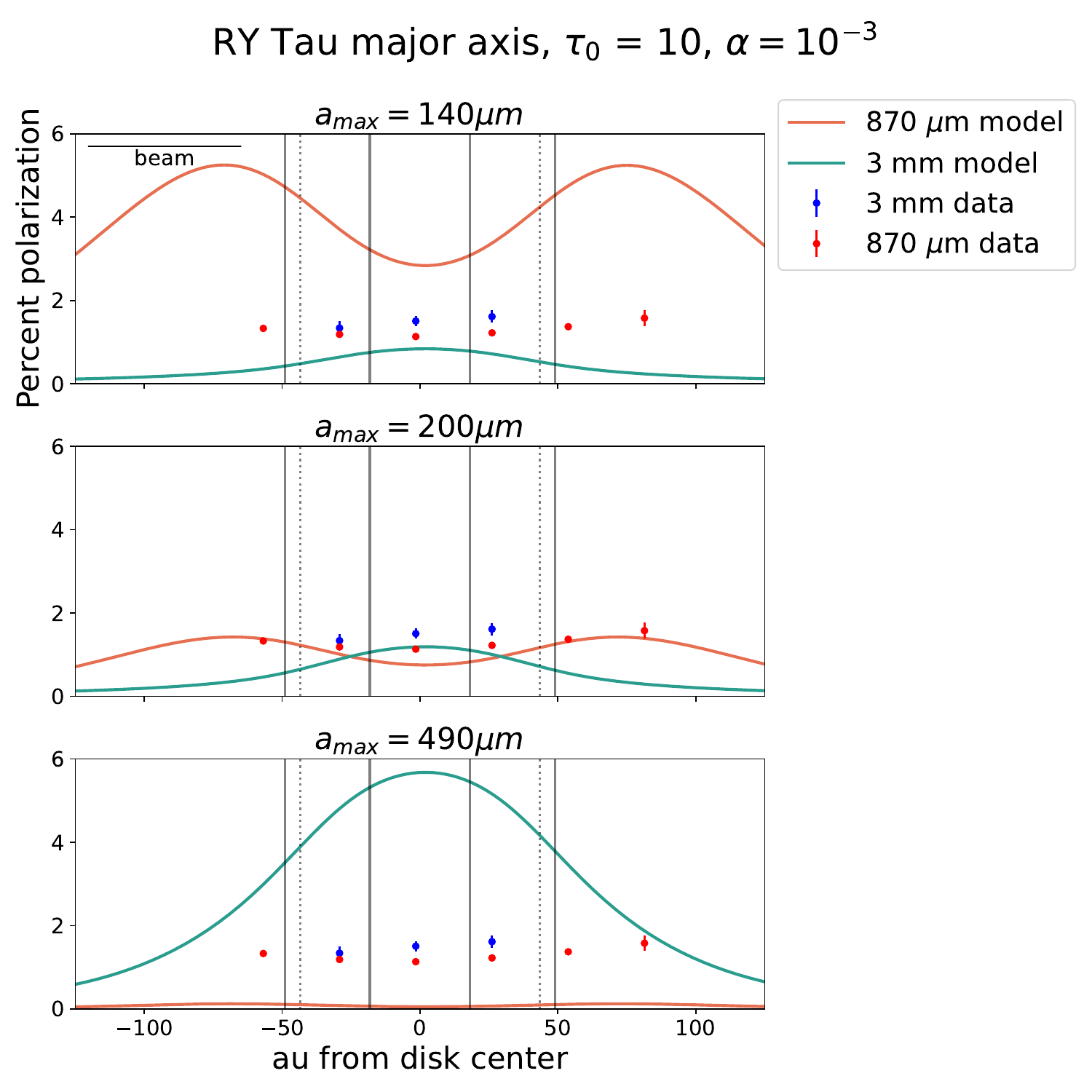}{0.48\textwidth}{(d)}}
\gridline{\fig{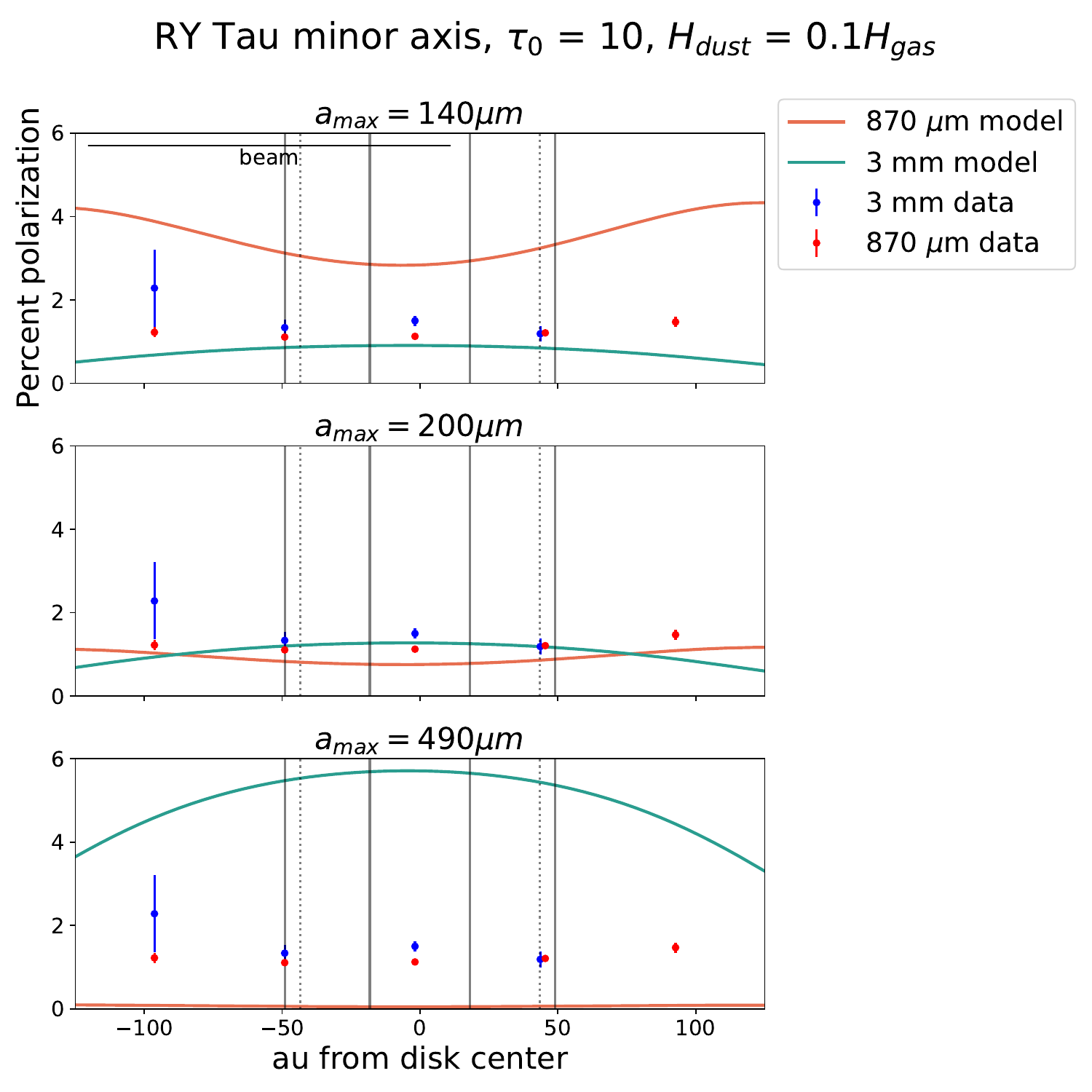}{0.48\textwidth}{(c)}
          \fig{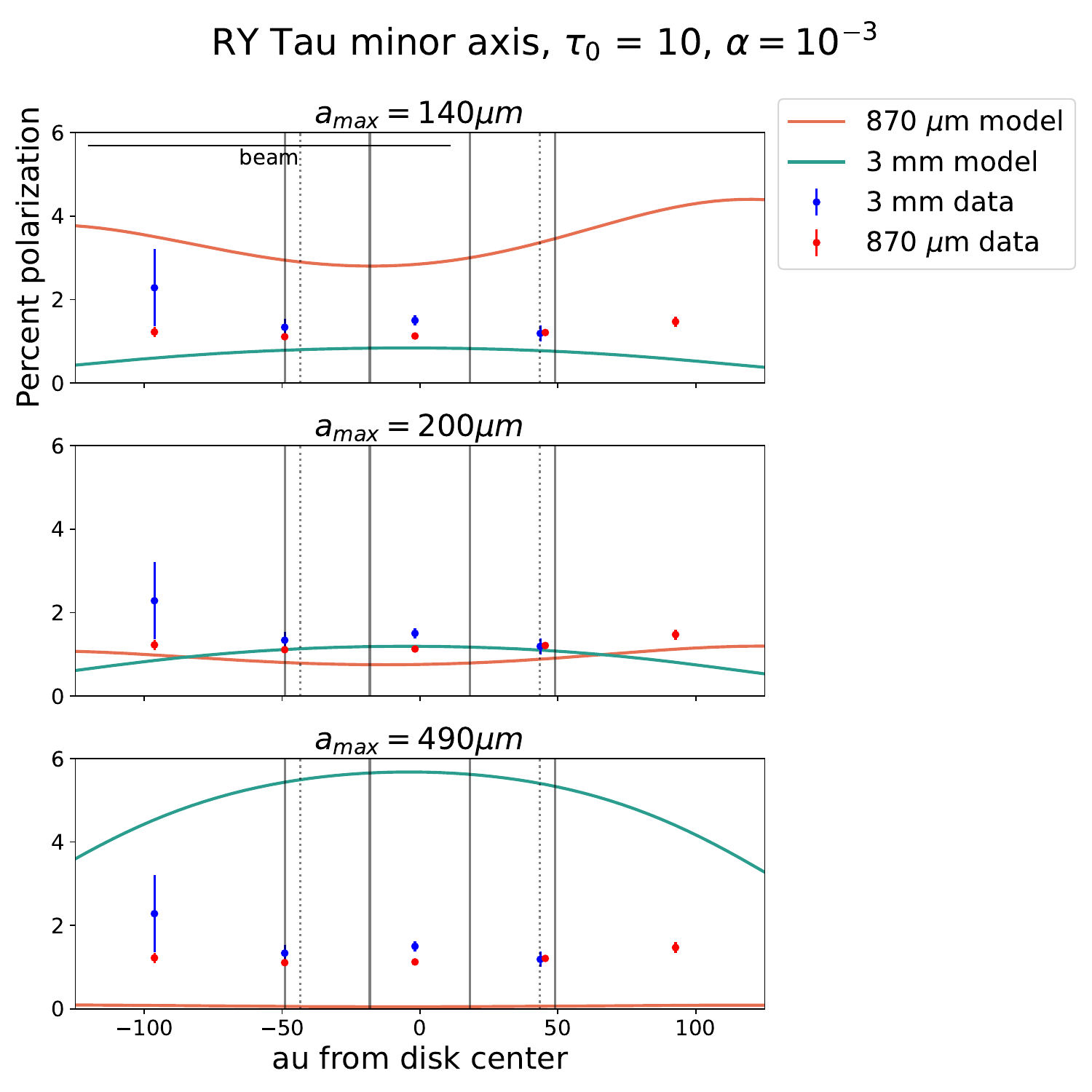}{0.48\textwidth}{(d)}}

\caption{Data vs. $\tau_0=10$ single-grain scattering models along the major and minor axes of RY Tau. Models have been convolved with a 0$\farcs$3 beam. Solid gray lines represent the locations of rings, and dashed gray lines represent the locations of gaps from \citet{2018ApJ...869...17L}. The beam scale bar represents the beam major axis of the 3 mm data.}
\label{singlegrain_rytau_tau10}
\end{figure*}

\begin{figure*}[ht]
\gridline{\fig{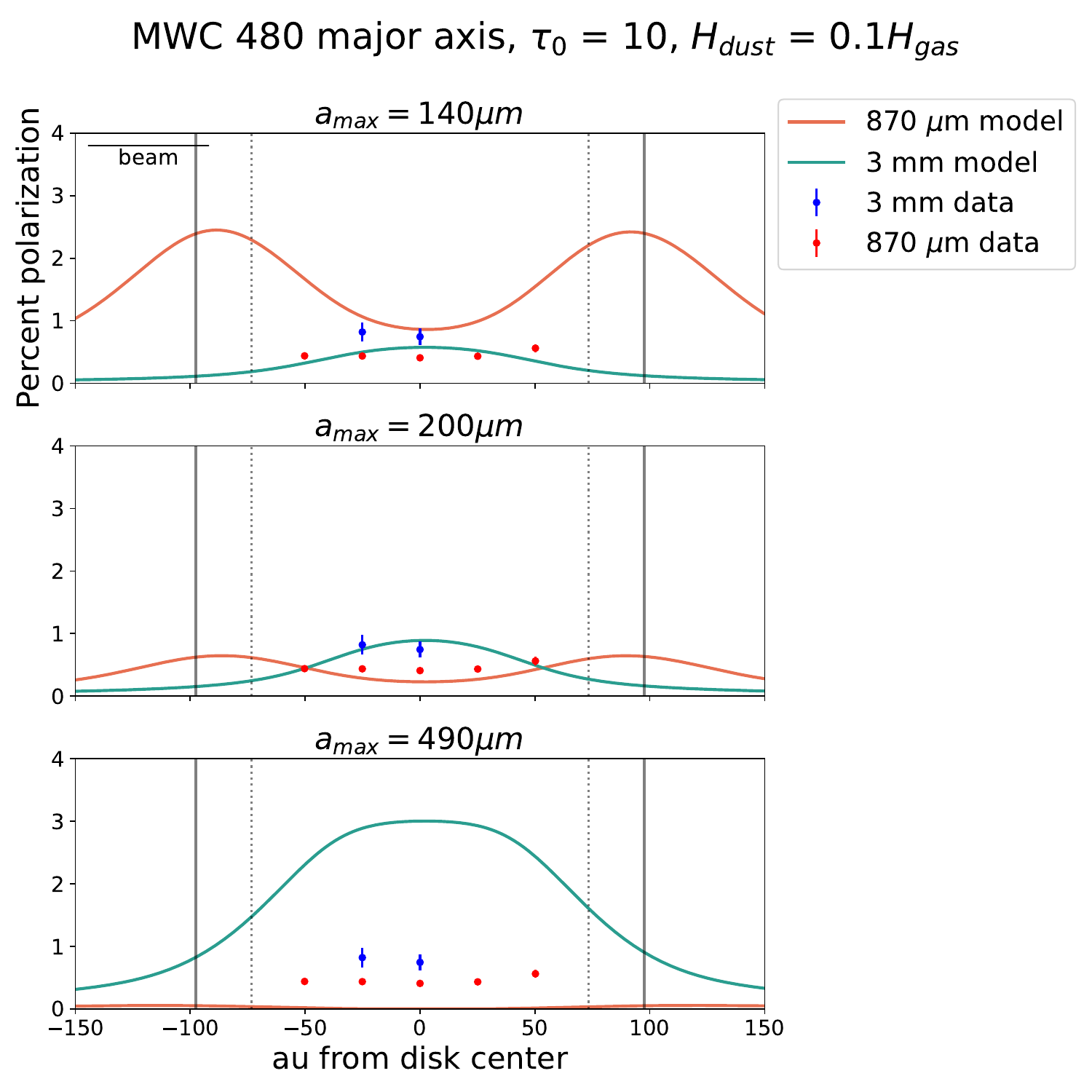}{0.48\textwidth}{(a)}
          \fig{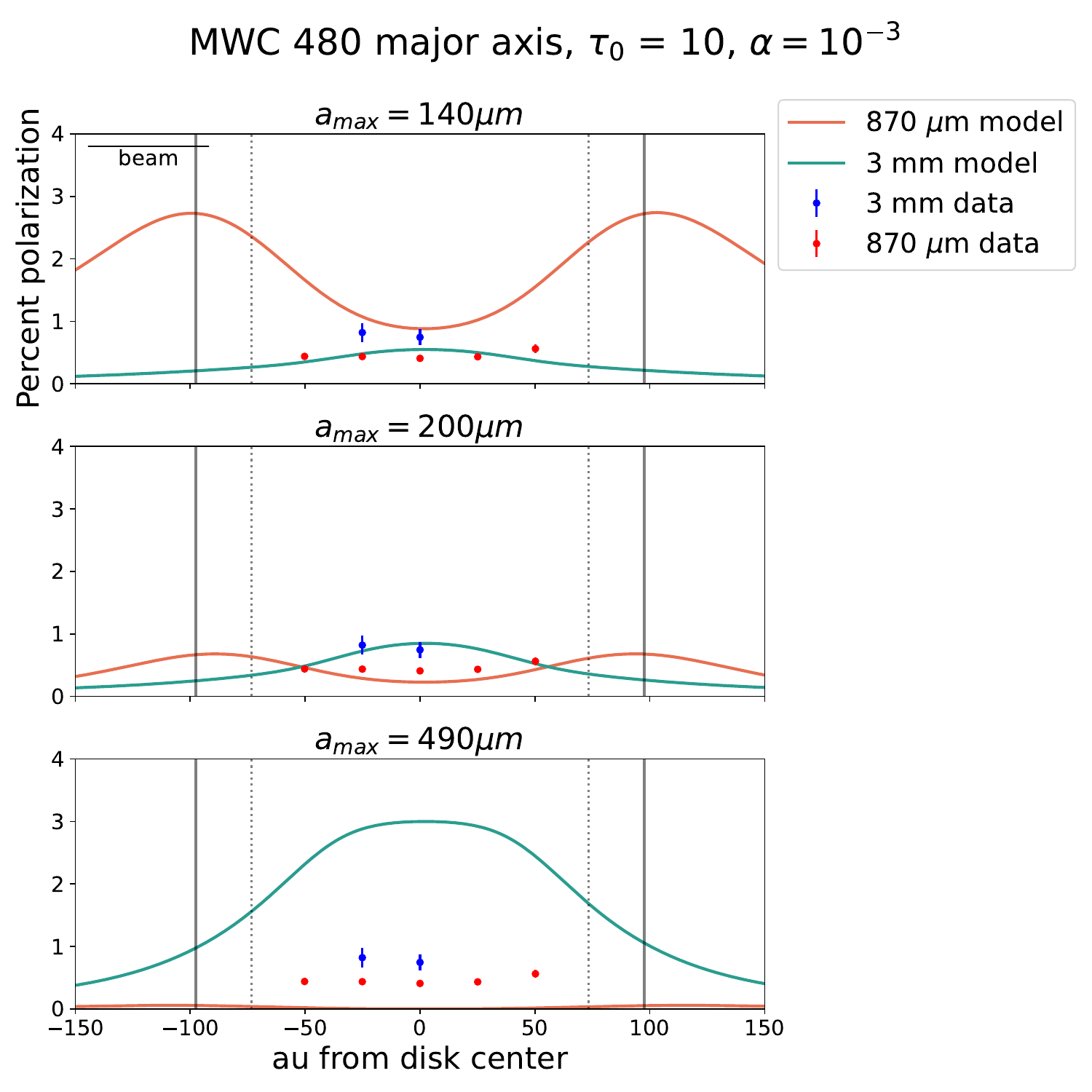}{0.48\textwidth}{(b)}}
\gridline{\fig{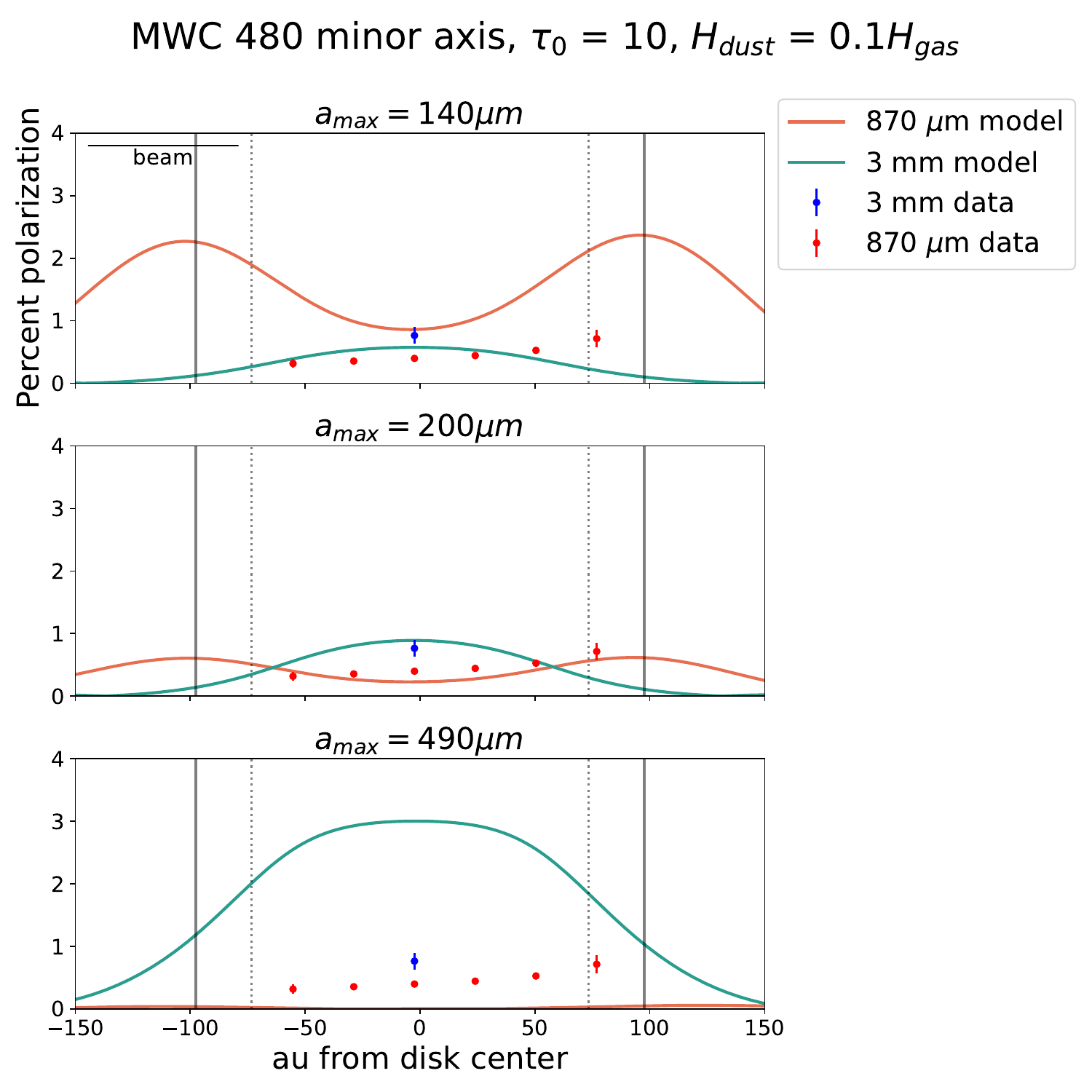}{0.48\textwidth}{(c)}
          \fig{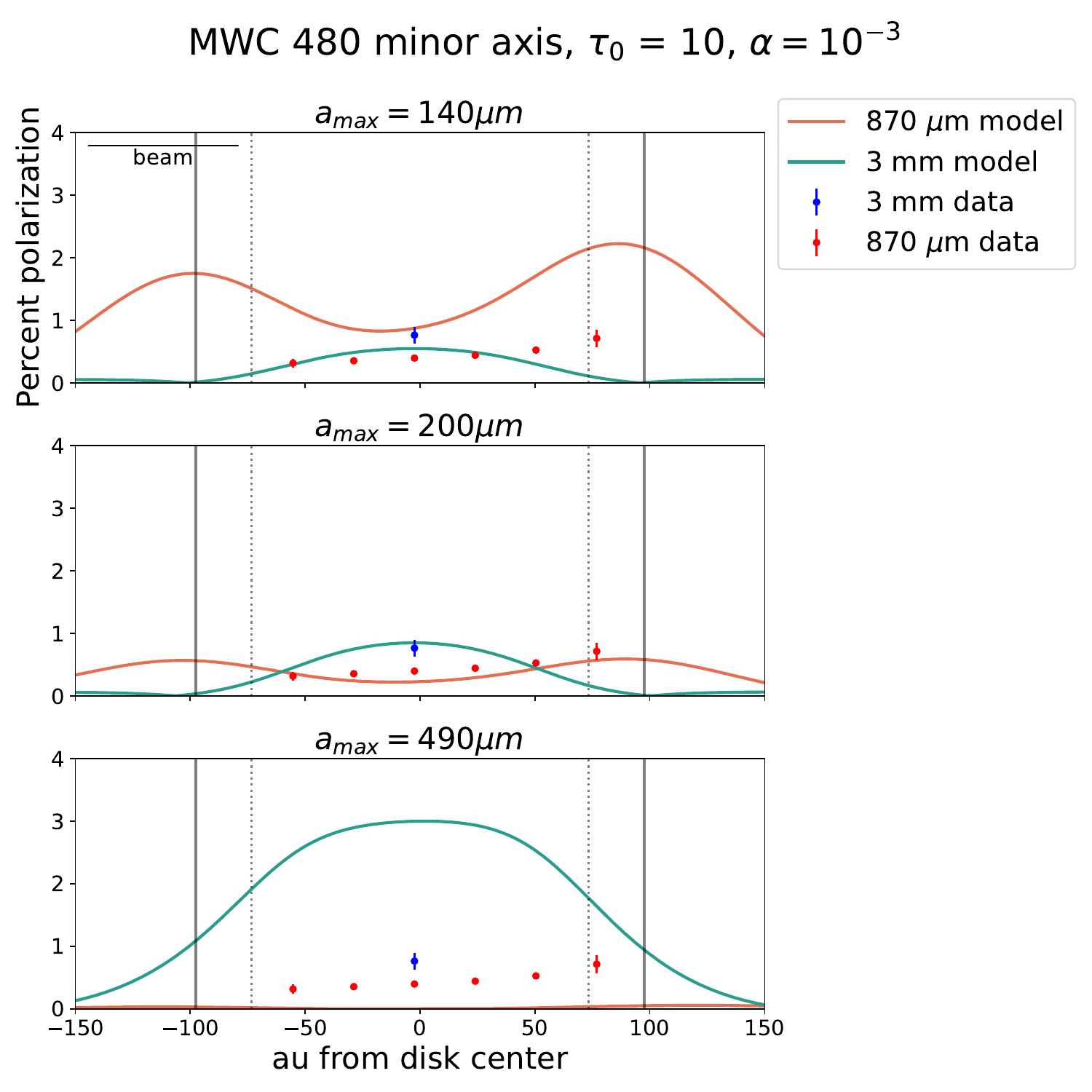}{0.48\textwidth}{(d)}}

\caption{Data vs. $\tau_0=10$ single-grain scattering models along the major and minor axes of MWC 480. Models have been convolved with a 0$\farcs$3 beam. Solid gray lines represent the locations of rings, and dashed gray lines represent the locations of gaps from \citet{2018ApJ...869...17L}. The beam scale bar represents the beam major axis of the 3 mm data.}
\label{singlegrain_mwc_tau10}
\end{figure*}

At high characteristic optical depths ($\tau_0=10$), the 140 $\mu$m dust can produces polarization fractions of up to $\sim$1\% in RY Tau and up to $\sim$0.5\% in MWC 480 (see Figures \ref{singlegrain_rytau_tau10} and \ref{singlegrain_mwc_tau10}). At high optical depths, the incident radiation field on a dust grain becomes increasingly isotropic, which attenuates the observed polarization fraction \citep{2017MNRAS.472..373Y}. This effect can be observed in the dip in polarization fraction at 870 $\mu$m at the center of the disk for models with high characteristic optical depths (see the top and middle panels of Figures \ref{singlegrain_rytau_tau10} and \ref{singlegrain_mwc_tau10}). The dip in polarization fraction in the center of the disk is present in the 870 $\mu$m data, which indicates that high optical depth models should be explored. The high optical depth models with $a_{max}$ = 490 $\mu$m produce a higher polarization fraction at 3 mm than at 870 $\mu$m in the center of the disk, but they produce a very low polarization fraction at the shorter wavelength see the bottom panels of \ref{singlegrain_rytau_tau10} and \ref{singlegrain_mwc_tau10}). The high optical depth models with $a_{max}$ = 200 $\mu$m produce a significant polarization fraction at both wavelengths, and have a higher polarization fraction at 3 mm than at 870 $\mu$m in the center of the disk see the middle panels of Figures \ref{singlegrain_rytau_tau10} and \ref{singlegrain_mwc_tau10}). These comparisons of the single-population, high optical depth models to the data indicate that future work should explore dust populations with maximum sizes between those with size parameters of 1 at 3 mm and 870 $\mu$m. 

\begin{figure*}[ht]
\gridline{\fig{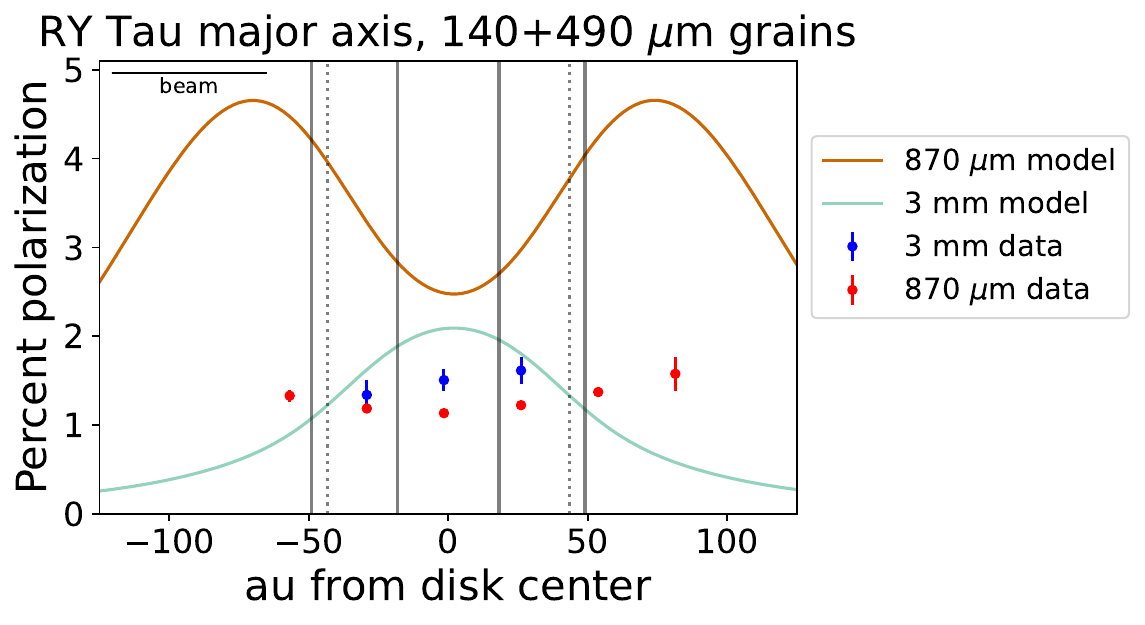}{0.48\textwidth}{(a)}
          \fig{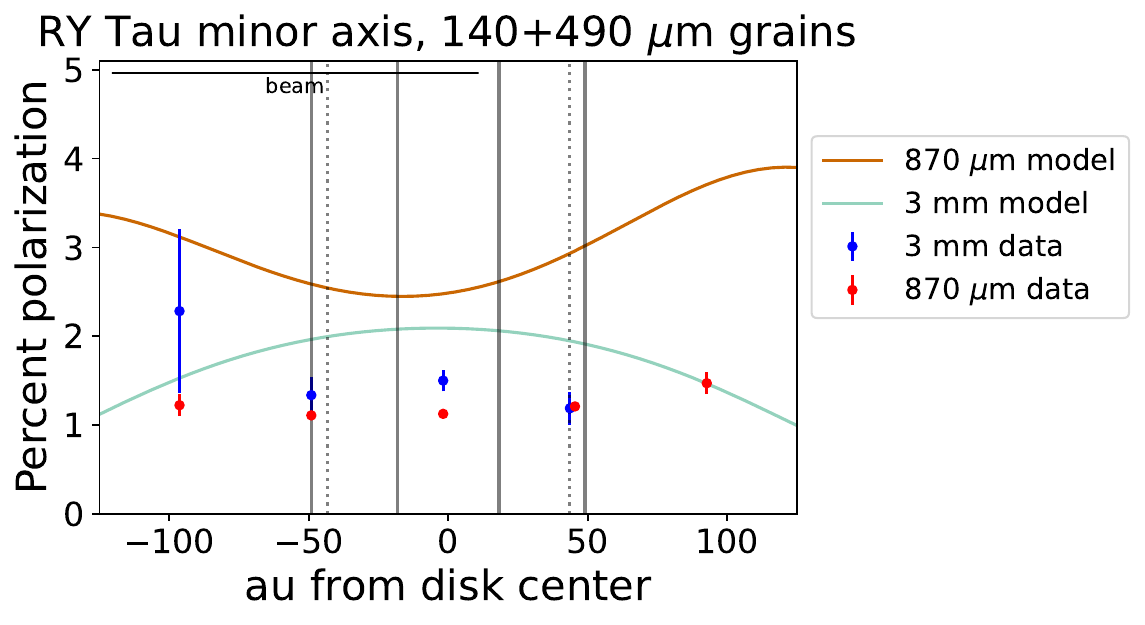}{0.48\textwidth}{(b)}}
\gridline{\fig{MWC_480_multigrain_majax_new.pdf}{0.48\textwidth}{(c)}
          \fig{MWC_480_multigrain_minax_new.pdf}{0.48\textwidth}{(d)}}

\caption{Data vs. models including two dust populations with maximum grain radii of 140 and 490 $\mu$m along the major and minor axes of MWC 480 and RY Tau. Models have been convolved with a 0$\farcs$3 beam. Solid gray lines represent the locations of rings, and dashed gray lines represent the locations of gaps from \citet{2018ApJ...869...17L}. The beam scale bar represents the beam major axis of the 3 mm data.}
\label{twograin_model}
\end{figure*}

Theoretical work \citep[e.g.,][]{Ueda2021} and observations \citep[e.g.,][]{Lee2017} both indicate that size-dependent vertical dust settling occurs in protoplanetary disks. This vertical stratification of dust by size would affect scattering polarization, as described in Section \ref{sec:discussion}. As a proof of concept, we present models of the scattering polarization resulting from 490 $\mu$m dust near the disk midplane and 140 $\mu$m dust above and below the midplane. In a disk with the inclination of RY Tau, this dust population produces a polarization fraction that would be observable at both 870 $\mu$m and 3 mm. However, the polarization fraction would be higher at 870 $\mu$m in the center of the disk, which is the opposite of the pattern seen in the data see Figure \ref{twograin_model} (a) and (b). For a disk with the inclination of MWC 480, this dust population leads to higher polarization fraction at 3 mm in the center of the disk (see Figure \ref{twograin_model} (c) and (d)).

\end{document}